\documentclass[10pt, superscriptaddress, aps, prd, amsmath, amssymb, nofootinbib]{revtex4-1}
\usepackage{amsfonts}
\usepackage{graphicx}
\usepackage{hyperref}

\def\wt#1{\widetilde{#1}}
\def\eq#1{{Eq.~(\ref{#1})}}
\def\fig#1{{Fig.~\ref{#1}}}
\def\bo#1{{\boldsymbol#1}}

\newcommand{\ud}{\mathrm{d}}

\begin{document}
\title{The suppressions of dijet azimuthal correlations in the future EIC}

\author{Ye-Yin Zhao}
\email{yeyin.zhao@gmail.com}
\affiliation{Key Laboratory of Quark and Lepton Physics (MOE) and Institute of Particle Physics, Central China Normal University (CCNU), Wuhan 430079, China}

\author{Ming-Mei Xu}
\affiliation{Key Laboratory of Quark and Lepton Physics (MOE) and Institute of Particle Physics, Central China Normal University (CCNU), Wuhan 430079, China}

\author{Li-Zhu Chen}
\affiliation{Nanjing University of Information Science and Technology, Nanjing 210044, China}

\author{Dong-Hai Zhang}
\affiliation{Key Laboratory of Quark and Lepton Physics (MOE) and Institute of Particle Physics, Central China Normal University (CCNU), Wuhan 430079, China}

\author{Yuan-Fang Wu}
\email{wuyf@mail.ccnu.edu.cn}
\affiliation{Key Laboratory of Quark and Lepton Physics (MOE) and Institute of Particle Physics, Central China Normal University (CCNU), Wuhan 430079, China}

\begin{abstract}
Quark-antiquark pair (or dijet) production at the electron-ion collider (EIC) has been argued to be one of most important processes that allowing to access the Weizs\"acker-Williams (WW) gluon distributions at small $x$ limit. Within the framework of Color Glass Condensate (CGC) effective field theory (EFT), we calculated the dijet cross sections and the azimuthal correlations by including the Sudakov resummations, numerical results shown that the back-to-back correlations are significantly suppressed when the Sudakov resummations are taken into account. In addition, by using the solutions of running-coupling Balitsky-Kovchegov (rcBK) equation, the unpolarized and linearly polarized WW gluon distributions both in coordinate and momentum space are presented.
\end{abstract}

\maketitle

\section{Introduction}\label{sec:intro}
%==============================================================================
Exploring the multi-dimensional structures and the detailed dynamics of a hadron has been one of the most primary goals in hadronic physics. In recent years, the so-called transverse momentum distributions (TMDs) associated with their evolutions, representing essential aspects of the partonic structures in momentum space (parton distributions not only as a function of longitudinal momentum fraction $x$, but also involves the intrinsic transverse momentum $k_\perp$), have received a lot of concerns both in theory and experiment sides. A future planed deep inelastic scattering (DIS) facilities, such as electron-ion colliders (EIC)~\cite{Boer:2011fh,Accardi:2012qut,Aschenauer:2017jsk}, LHeC~\cite{AbelleiraFernandez:2012cc,Agostini:2020fmq} and EicC~\cite{Anderle:2021wcy}, will provide the cleanest environment to determine these fundamental quantities over a broad range of kinematics at high accuracy.

The quark-antiquark pair (or dijet) production at DIS has been argued to be an important process that allowing to access the partonic structures in momentum space. Especially, when the produced jets are mostly back-to-back, the novel process can be used to as a proxy to probe the Weizs\"acker-Williams (WW) gluon distribution~\cite{Dominguez:2011wm}. The WW gluon distribution, as one type of gluon TMDs, which features an intrinsic gauge link structures: either future-pointing or past-pointing gauge links in light-cone gauge and can be interpreted as the number density of gluons inside the scattered target~\cite{Kovchegov:1998bi,McLerran:1998nk,Dominguez:2011wm}. As addressed in Ref.~\cite{Mulders:2000sh}, a nontrivial feature has been realized that these gluons can be linearly polarized even if the hadron is unpolarized, especially, if sizable, can give a large impact on physical observables, for examples, the (pseudo)scalar particle productions in hadronic collisions~\cite{Sun:2011iw,Boer:2011kf,Echevarria:2015uaa}, as well as the azimuthal correlations of dijets in DIS~\cite{Dominguez:2010xd,Dominguez:2011wm,Boer:2010zf,Pisano:2013cya,Zheng:2014vka,Boer:2016fqd,Efremov:2017iwh,Efremov:2018myn,Dumitru:2015gaa,Dumitru:2018kuw}. In Refs.~\cite{Metz:2011wb,Dominguez:2011br}, for dijets production in DIS, it has been shown that unpolarized WW gluon distribution associated with its partner, linearly polarized one, enter the cross section through an azimuthal modulation. Recent developments also shown that the two gluon distributions not only exist in DIS processes, but also wildly appear in other processes, such as the dijets or quark-antiquark pairs~\cite{Boer:2009nc,Akcakaya:2012si,Marquet:2017xwy}, photon pairs~\cite{Qiu:2011ai}, muon pairs~\cite{Pisano:2013cya}, quarkonium~\cite{Boer:2016bfj,Lansberg:2017dzg,Lansberg:2017tlc,Mukherjee:2016cjw} productions in hadronic collisions. A particular feature is that the unpolarized WW gluon distributions are always accompanied by its linearly polarized parts through an azimuthal modulation, this suggests that these gluon distributions (especially for the linearly polarized parts) can be extracted by measuring the azimuthal observables. Another interesting feature has been pointed out that the linearly polarized WW gluon distribution will tend to its saturated bound at large transverse momenta, while is suppressed at small transverse momenta region~\cite{Metz:2011wb,Dominguez:2011br}, which was then further confirmed by numerical calculations in Refs.~\cite{Dumitru:2015gaa,Dumitru:2018kuw}, for recent review please see Ref.~\cite{Petreska:2018cbf} and therein. Up to now, the two gluon distributions, apart from the positive bound of linearly polarized part, are still little known experimentally, therefore, to precisely extract and quantitatively understand the behaviors of the two gluon distributions at high energy limit, which allows us in depth understanding the transverse tomography of proton and nucleus, as well as the underlying dynamics at high energy scatterings.

In this paper, we continue to concentrate on the quark-antiquark pair (or dijet) production in a special process that a virtual photon scattering off a hadronic target, this process typically corresponds to the quark-antiquark (or dijet) production in DIS. Due to a broad range of kinematics that can be accessed in the future EIC, this process allows to mapping out the evolutions of gluon TMDs. Especially, when the scattered energy is sufficiently large (or equivalent to small-$x$), a high density regime can be probed, in this sense, gluon saturation effects will become more predominant, and hence the behaviors of gluon TMDs in the dense regime can be investigated as well. In the correlation limit, prior works have shown that this process can be consistently described in the framework of Color Glass Condensate (CGC) theory, and simultaneously, the relevant two WW gluon distributions can be expressed in terms of Wilson line correlators that making them can be calculated by using the solutions of Jalilian-Marian-Iancu-McLerran-Weigert-Leonidov-Kovner (JIMWLK) equation~\cite{Metz:2011wb,Dominguez:2011br,Dumitru:2015gaa,Dumitru:2018kuw}. However, the special limit has a strong restriction on the dijets, which requires the transverse momentum imbalance $q_\perp$ is much smaller than the relative transverse momentum $P_\perp$, in this case, there is another class of effects, known as Sudakov resummations, can give sizable contribution to the cross section, and this should be taken into account when exploring the behaviors of the two gluon distributions in the considered process. Other phenomenological studies have shown that such resummations is very important in quantitatively describing experimental observables~\cite{Qiu:2000ga,Hautmann:2008vd,Bozzi:2010xn,Deak:2011ga,Cieri:2015rqa,Sun:2015doa,Mueller:2016gko,Stasto:2018rci,Banfi:2008qs,Mueller:2012uf,Mueller:2013wwa,Sun:2014gfa,Mueller:2016xoc}. Therefore, in the present work, we plan to conduct this direction by including the Sudakov resummations, and show how the resummations affect the azimuthal correlations of the dijets. Furthermore, as shown in Refs.~\cite{Dumitru:2015gaa,Dumitru:2018kuw,Marquet:2016cgx,Marquet:2017xwy}, thanks to the numerical implementation of JIMWLK equation, the two gluon distributions appeared in the considered process can be obtained. However, the numerical implementation is only achieved by fixing coupling constant. The running-coupling corrections, as an important QCD dynamics, has not been performed yet in their studies. In the present work, we prefer to work within the Gaussian approximation of JIMWLK evolution and use the running-coupling Balitsky-Kovchegov (rcBK) equation to take the relevant calculations, because we expect the running-coupling corrections here to be much more important than that of Gaussian approximation. Besides that, we stress the rcBK solutions are well constrained from DIS data~\cite{Albacete:2010sy} and have been received a great successful in describing particle spectrum at hadronic collisions~\cite{Albacete:2010bs,Zhao:2013kea}, therefore, one can be firmly convinced that the gluon TMDs can be well studied by using the solutions of rcBK equation.

The paper is organized as follows. In next section, we will begin with a brief review of the CGC results for the quark-antiquark pair production in DIS, and show that the cross sections are the sensitive probes of gluon TMDs (unpolarized and linearly polarized WW gluon distributions) inside the scattered target at small-$x$. In Sec.~\ref{sec:ww}, we discuss the detailed behaviors of unpolarized and linearly polarized WW gluon distributions in the saturation model and then present the numerical results by using the solutions of rcBK equation. In Sec.~\ref{sec:results}, we give our numerical results for the dijet cross sections, as well as discuss how the effects of Sudakov resummations and gluon saturation affect their azimuthal correlations. Finally, a short summary is given in Sec.~\ref{sec:summary}.

\section{DIS dijet production in color glass condensate formalism}\label{sec:Xsection}
%==============================================================================
In this section, we firstly give a brief review of the quark-antiquark pair (or dijet) differential cross section in the CGC formalism, and show that, by applying the ``back-to-back correlation limit'', the cross sections are expressed in terms of WW gluon distributions.

To begin with, we consider a high energetic virtual photon $\gamma^\ast$ with four-momentum $p$ scattering off a hadronic target, after that, a quark-antiquark pair $q\bar{q}$ is produced. This process typically corresponds to quark-antiquark pair (or dijet\footnote{Strictly speaking, for dijet production, the chanel $\gamma^\ast q\to\gamma^\ast qg$ should be included. However, since we are only interest in small-$x$ regime, in which the gluon density is very large as compared to the quarks, in this sense, the contributions of quarks can be neglected. Otherwise, the quark part enters and gives an additional contribution to the cross section~\cite{Pisano:2013cya,Kovchegov:2015zha}.}) production in DIS via the exchange of a virtual photon. For this process, the hadronic target can be regarded as classical gluon field due to the dominance of gluon saturation effects~\cite{Gribov:1984tu,Mueller:1985wy,McLerran:1993ka,McLerran:1993ni,McLerran:1994vd}. Therefore, it is convenient to calculate the cross section in the CGC formalism, the standard way is to consider the virtual photon $\gamma^\ast$ fluctuates into a quark-antiquark pair $q\bar{q}$, and then traverse the background field in an eikonal way, simultaneously, the fluctuated quarks receive a contribution to the scattering amplitude with a Wilson line. After averaging over the photon's polarization and summing over the quark's helicities and colors, the differential cross section for this process can be expressed as follow~\cite{Metz:2011wb,Dominguez:2011wm},
\begin{equation}\label{eq:1}
\begin{split}
\frac{\ud \sigma^{\gamma^\ast_{T,L}A\to q\bar{q}X} }{\ud^3k_1 \ud^3k_2}=
& N_c \alpha_{\text{em}} e^2_q \delta(p^+ - k^+_1 - k^+_2) \int
\frac{\ud^2\bo{x}_1}{(2\pi)^2} \frac{\ud^2\bo{x}'_1}{(2\pi)^2}
\frac{\ud^2\bo{x}_2}{(2\pi)^2} \frac{\ud^2\bo{x}'_2}{(2\pi)^2} \\
& \times e^{-i\bo{k}_{1\perp} \cdot (\bo{x}_1 - \bo{x}'_1)} e^{-i\bo{k}_{2\perp} \cdot (\bo{x}_2 - \bo{x}'_2)}
\sum_{\lambda\alpha\beta} \psi^{T,L\lambda}_{\alpha\beta}(\bo{x}_1 - \bo{x}_2) \psi^{T,L\lambda\ast}_{\alpha\beta}(\bo{x}'_1 - \bo{x}'_2) \\
& \times \left[1 + S^{(4)}_{x_g}(\bo{x}_1,\bo{x}_2;\bo{x}'_2,\bo{x}'_1) - S^{(2)}_{x_g}(\bo{x}_1,\bo{x}_2) - S^{(2)}_{x_g}(\bo{x}'_2,\bo{x}'_1)\right]
\end{split}
\end{equation}
where $\alpha_{\text{em}}$ is the electromagnetic structure constant, $N_c$ and $e_q$ are the number of colors and the electrical charge of the produced quark. The two-dimensional variables $\bo{x}_1$ and $\bo{x}_2$ denote the transverse coordinate of the outgoing quark and antiquark in the amplitude respectively. Clearly, the variable $\bo{u}=\bo{x}_1-\bo{x}_2$ and $\bo{v}=z\bo{x}_1+(1-z)\bo{x}_2$ are the transverse size of the produced quark-antiquark pair and the transverse position of the incoming virtual photon, where $z=k^+_1/p^+$ is the longitudinal momentum fraction of the outgoing quark with respect to the incoming virtual photon. Similar notations hold for the complex conjugate amplitude. The splitting function $\psi^{T,L\lambda}_{\alpha\beta}(p^+,z,\bo{u})$ denotes the virtual photon $\gamma^\ast$ with longitudinal momentum $p^+$ and virtuality $Q^2$ fluctuates into a quark-antiquark pair with size $\bo{u}$, this function can be calculated in light-cone perturbative theory and be expressed in the following form~\cite{Kowalski:2006hc},
\begin{align}
\psi^{T\lambda}_{\alpha\beta}(p^+,z,\bo{u}) &= 2\pi\sqrt{\frac{2}{p^+}}\left\{
\begin{array}{rl}
i\epsilon_f K_1(\epsilon_f \vert \bo{u}\vert)\frac{\bo{u}\cdot\epsilon^{(1)}_\perp}{\vert\bo{u}\vert}\left[\delta_{\alpha +}\delta_{\beta +}(1-z) + \delta_{\alpha -}\delta_{\beta -}z\right] + \delta_{\alpha -}\delta_{\beta +} m_q K_0(\epsilon_f\vert \bo{u}\vert), & \lambda=1, \\
i\epsilon_f K_1(\epsilon_f \vert \bo{u}\vert)\frac{\bo{u}\cdot\epsilon^{(2)}_\perp}{\vert\bo{u}\vert}\left[\delta_{\alpha -}\delta_{\beta -}(1-z) + \delta_{\alpha +}\delta_{\beta +}z\right] + \delta_{\alpha +}\delta_{\beta -} m_q K_0(\epsilon_f\vert \bo{u}\vert), & \lambda=2,
\end{array}
\right. \label{eq:2}\\
\psi^{L\lambda}_{\alpha\beta}(p^+,z,\bo{u}) &= 2\pi\sqrt{\frac{4}{p^+}}z(1-z)Q K_0(\epsilon_f\vert \bo{u}\vert)\delta_{\alpha\beta}.\label{eq:3}
\end{align}
where $\alpha, \beta$ are the helicities of produced quark and antiquark, $\lambda$ is the photon's polarization, $T,L$ denote the transverse and longitudinal components of the wavefunction, $\epsilon^2_f=z(1-z)Q^2 + m^2_q$ with $m_q$ is the mass of  produced quark.

It should be noted that the interactions between the produced quark-antiquark pair and the scattered target are fully characterized by Wilson line correlators, as shown in the last line of \eq{eq:1}. Specifically, the term equal to 1 means the virtual photon freely pass through the target, and the quark-antiquark pair without scattering in the amplitude and complex conjugate amplitude. The interference terms $S^{2}_{x_g}(\bo{x}_1,\bo{x}_2)$ and $S^{(2)}_{x_g}(\bo{x}'_2,\bo{x}'_1)$, known as dipole, denote the quark-antiquark pair scatters off the target either in the amplitude or the complex conjugate amplitude. The four-point function $S^{(4)}_{x_g}(\bo{x}_1,\bo{x}_2,\bo{x}'_1,\bo{x}'_2)$, named as quadrupole, describes the quark-antiquark pair undergoes multi-scattering with target both in amplitude and complex conjugate amplitude. The dipole and quadrupole are defined as:
\begin{eqnarray}
S^{(2)}_{x_g}(\bo{x}_1,\bo{x}_2) &=& \frac{1}{N_c}\left\langle\text{Tr}U(\bo{x}_1)U^\dagger(\bo{x}_2)\right\rangle_{x_g}, \\
S^{(4)}_{x_g}(\bo{x}_1,\bo{x}_2;\bo{x}'_2,\bo{x}'_1) &=& \frac{1}{N_c}\left\langle\text{Tr}U(\bo{x}_1)U^\dagger(\bo{x}'_1)U(\bo{x}'_2)U^\dagger(\bo{x}_2)\right\rangle_{x_g}, \\
\text{with} \quad U(\bo{x}) &=& \mathcal{P}\exp\left[ig_s\int\ud x^+ T^c A^-_c(x^+, \bo{x})\right].
\end{eqnarray}
The notation $\langle\cdots\rangle_{x_g}$ used here denotes the averaging over all possible distributions of color sources and $x_g$ being the fraction of longitudinal momentum in the scattered target.

As shown in the above, the quark-antiquark pair (or dijet) production in DIS process can be quite well described in the CGC formalism, and the cross section directly probes the internal dynamics of the scattered target. It should be reminded that the CGC approach does not dependent on momentum scales and, in principle, can be applied in any scattering process when high density or ``saturation'' regime accessed. This treatment is different from the TMD factorization approach, since the later one is strongly dependent on momentum scales. In order to recover the TMD results, it is useful to change variables as follow,
\begin{equation}
\bo{x}_1 = \bo{v} + (1-z)\bo{u} \quad\text{and}\quad \bo{x}_2 = \bo{v} - z\bo{u}
\end{equation}
then, \eq{eq:1} casts into,
\begin{equation}\label{eq:8}
\begin{split}
\frac{\ud\sigma^{\gamma^\ast_{T,L}A\to q\bar{q}X}}{\ud^3k_1\ud^3k_2} =& N_c\alpha_{\text{em}}e^2_q\delta(q^+-k^+_1-k^+_2)\int\frac{\ud^2\bo{u}}{(2\pi)^2}\frac{\ud^2\bo{u}'}{(2\pi)^2}\frac{\ud^2\bo{v}}{(2\pi)^2}\frac{\ud^2\bo{v}'}{(2\pi)^2}\\
&\times e^{-iq_\perp\cdot(\bo{v}-\bo{v}')} e^{-iP_\perp\cdot(\bo{u}-\bo{u}')}\sum_{\lambda\alpha\beta}\psi^{L,T\lambda}_{\alpha\beta}(\bo{u})\psi^{L,T\lambda\ast}_{\alpha\beta}(\bo{u}')\\
&\times \left[1+\frac{1}{N_c}\left\langle\text{Tr}U(\bo{v}+(1-z)\bo{u})U^\dagger(\bo{v}'+(1-z)\bo{u}')U(\bo{v}'-z\bo{u}')U^\dagger(\bo{v}-z\bo{u})\right\rangle_{x_g}\right.\\
&\quad\quad\left.-\frac{1}{N_c}\left\langle\text{Tr}U(\bo{v}+(1-z)\bo{u})U^\dagger(\bo{v}-z\bo{u})\right\rangle_{x_g}-\frac{1}{N_c}\left\langle\text{Tr}U(\bo{v}'-z\bo{u}')U^\dagger(\bo{v}'+(1-z)\bo{u}')\right\rangle_{x_g}\right]
\end{split}
\end{equation}
where $q_\perp=k_{1\perp}+k_{2\perp}$ and $P_\perp=(1-z)k_{1\perp}-zk_{2\perp}$ are the transverse momentum imbalance and the relative momentum of the quark pair. Obviously, the two variables are the conjugate momenta respect to $\bo{v}$ and $\bo{u}$. In what follows, we are only interest in a special kinematic limit\footnote{Beyond the correlation limit, this process has been studied in Ref.~\cite{Mantysaari:2019hkq}, which gives insight into the dynamics of multi-point Wilson line correlators at small $x$ limit.}, also known as the ``back-to-back correlation limit''~\cite{Dominguez:2011wm}, in which the transverse momentum imbalance is much smaller than the relative momentum, i.e. $\vert q_\perp\vert\ll\vert P_\perp\vert$. With this limitation, the corresponding variables $\bo{u}$ and $\bo{u}'$ are much smaller than $\bo{v}$ and $\bo{v}'$, in this sense, terms involving multi-point Wilson line correlators can be expanded with Taylor series in coordinate space. To leading order in $\bo{u}$ and $\bo{u}'$\footnote{Following this procedure, higher order can be obtained, next-to-leading power order can be found in Ref.~\cite{Dumitru:2016jku}.}, \eq{eq:8} can be rewritten as,
\begin{equation}\label{eq:9}
\begin{split}
\frac{\ud\sigma^{\gamma^\ast_{T,L}A\to q\bar{q}X}}{\ud^3k_1\ud^3k_2}=&N_c\alpha_{\text{em}}e^2_q\delta(q^+-k^+_1-k^+_2)\int\frac{\ud^2\bo{u}}{(2\pi)^2}\frac{\ud^2\bo{u}'}{(2\pi)^2}\frac{\ud^2\bo{v}}{(2\pi)^2}\frac{\ud^2\bo{v}'}{(2\pi)^2}\\
&\times e^{-iq_\perp\cdot(\bo{v}-\bo{v}')}e^{-iP_\perp\cdot(\bo{u}-\bo{u}')}\sum_{\lambda\alpha\beta}\psi^{L,T\lambda}_{\alpha\beta}(\bo{u})\psi^{L,T\lambda\ast}_{\alpha\beta}(\bo{u}')\\
&\times\left[-\bo{u}_i\bo{u}'_j\frac{1}{N_c}\left\langle\text{Tr}\left[\partial^iU(\bo{v})\right]U^\dagger(\bo{v}')\left[\partial^jU(\bo{v}')\right]U^\dagger(\bo{v})\right\rangle_{x_g}\right].
\end{split}
\end{equation}
From the above expression, one can see that the differential cross section can be factorized into two parts, one is involved hard scale $P_\perp$, which is referred to as hard part, the left one is related to $q_\perp$, this is the quantity that directly probe the transverse momentum of incoming gluon from the hadronic target. Together with the saturation scale $Q_{sg}$, there are three scales in this scattering process. Notice that, although $\vert q_\perp\vert\ll \vert P_\perp\vert$, the value of $q_\perp$ can be order of (or softer than) the saturation scale $Q_{sg}$, in this case, gluon saturation effects still remain important and this process can indeed provide information on the internal dynamics in the dense regime.

Within the back-to-back correlation limit, due to the two separated scales, i.e. $\vert P_\perp\vert\gg\vert q_\perp\vert$, there is another class of effects with a double logarithmic form $\ln\frac{P^2_\perp}{q^2_\perp}$ can give large contribution to the cross section and this should be included. This factor, also known as Sudakov resummations, comes from the resummation of soft collinear gluon radiation of incoming gluons from the scattered target. In the considered process, on the cross section level, which can be obtained by calculating one-loop correction of the quark pair~\cite{Mueller:2012uf,Mueller:2013wwa}. After including this contribution, \eq{eq:9} becomes,
\begin{equation}\label{eq:10}
\begin{split}
\frac{\ud\sigma^{\gamma^\ast_{T,L}A\to q\bar{q}X}}{\ud^3k_1\ud^3k_2}=&N_c\alpha_{\text{em}}e^2_q\delta(q^+-k^+_1-k^+_2)\int\frac{\ud^2\bo{u}}{(2\pi)^2}\frac{\ud^2\bo{u}'}{(2\pi)^2}\frac{\ud^2\bo{v}}{(2\pi)^2}\frac{\ud^2\bo{v}'}{(2\pi)^2}\\
&\times e^{-iq_\perp\cdot(\bo{v}-\bo{v}')}e^{-iP_\perp\cdot(\bo{u}-\bo{u}')}\sum_{\lambda\alpha\beta}\psi^{L,T\lambda}_{\alpha\beta}(\bo{u})\psi^{L,T\lambda\ast}_{\alpha\beta}(\bo{u}')e^{-S_{\text{sud}}(P_\perp,\bo{v}-\bo{v}')}\\
&\times\left[-\bo{u}_i\bo{u}'_j\frac{1}{N_c}\left\langle\text{Tr}\left[\partial^iU(\bo{v})\right]U^\dagger(\bo{v}')\left[\partial^jU(\bo{v}')\right]U^\dagger(\bo{v})\right\rangle_{x_g}\right].
\end{split}
\end{equation}

To leading order, the Sudakov factor can be expressed as~\cite{Mueller:2013wwa}
\begin{equation}
S_{\text{sud}}(P_\perp,\bo{v}-\bo{v}')=\int^{P^2_\perp}_{c^2_0/(\bo{v}-\bo{v}')^2}\frac{\ud \mu^2}{\mu^2}\frac{N_c}{2}\frac{\alpha_s}{\pi}\ln\frac{P^2_\perp}{\mu^2}
\end{equation}
where $c_0=2e^{-\gamma_E}$ with $\gamma_E$ is the Euler constant.

It should be noted that the Sudakov factor only resums the soft radiations of incoming gluons from the scattered target, and this contribution is independent of $\bo{u}$ and $\bo{u}'$. Therefore, one can straightforwardly integrate over $\bo{u}$ and $\bo{u}'$ by collecting \eq{eq:2} and \eq{eq:3}, then \eq{eq:10} becomes,
\begin{align}
\frac{\ud\sigma^{\gamma^\ast_TA\to q\bar{q}X}}{\ud^3k_1\ud^3k_2}=&\alpha_\text{em} e^2_q\delta(x_{\gamma^\ast}-1)z(1-z)\left\lbrace 2\left[z^2+(1-z)^2\right]\left[\frac{\delta^{ij}}{(P^2_\perp + \epsilon^2_f)^2}-\frac{4\epsilon^2_f P^i_\perp P^j_\perp}{(P^2_\perp + \epsilon^2_f)^4}\right] + \frac{8m^2_qP^i_\perp P^j_\perp}{(P^2_\perp+\epsilon^2_f)^4}\right\rbrace\nonumber\\
&\times \int\frac{\ud^2\bo{v}}{(2\pi)^2}\frac{\ud^2\bo{v}'}{(2\pi)^2}e^{-S_\text{sud}(P_\perp,\bo{v}-\bo{v}')}e^{-iq_\perp\cdot(\bo{v}-\bo{v}')}\left[-\left\langle\text{Tr}\left[\partial^iU(\bo{v})\right]U^\dagger(\bo{v}')\left[\partial^jU(\bo{v}')\right]U^\dagger(\bo{v})\right\rangle_{x_g}\right],\\
\frac{\ud\sigma^{\gamma^\ast_L A\to q\bar{q}X}}{\ud^3 k_1\ud^3 k_2}=&\alpha_\text{em}e^2_q\delta(x_{\gamma^\ast}-1)z(1-z)\left[4z^2(1-z)^2Q^2\frac{4P^i_\perp P^j_\perp}{(P^2_\perp+\epsilon^2_f)^4}\right]\nonumber\\
&\times\int\frac{\ud^2\bo{v}}{(2\pi)^2}\frac{\ud^2\bo{v}'}{(2\pi)^2}e^{-S_\text{sud}(P_\perp,\bo{v}-\bo{v}')}e^{-iq_\perp\cdot(\bo{v}-\bo{v}')}\left[-\left\langle\text{Tr}\left[\partial^iU(\bo{v})\right]U^\dagger(\bo{v}')\left[\partial^jU(\bo{v}')\right]U^\dagger(\bo{v})\right\rangle_{x_g}\right].
\end{align}
Due to $P_\perp$ is the conjugate momentum respect to $\bo{u}$, the orientation of $\bo{u}_i$ holds in the above expressions, and is represented by $P^i_\perp$ after taking the Fourier transformation. Another piece is the double derivation of quadrupole, this one is related to the operator definition of WW gluon distribution. When averaging over impact parameter $B_\perp=\frac{1}{2}(\bo{v}+\bo{v}')$, it can be expressed as
\begin{equation}\label{eq:15}
xW^{ij}(x_g, \bo{v}-\bo{v}') = -\frac{2S_\perp}{(2\pi)^2\alpha_s}\left\langle\text{Tr}\left[\partial^i U(\bo{v})\right]U^\dagger(\bo{v}')\left[\partial^j U(\bo{v}')\right]U^\dagger(\bo{v})\right\rangle_{x_g}
\end{equation}
in which $S_\perp=\int \ud^2B_\perp$ is the transverse area of the scattered target. Generally speaking, $xW^{ij}$ is a $2\times 2$ matrix, it can be decomposed into various different tensor structures. In what follows, it is convenient to separate it as trace and traceless parts, for positive definite, the corresponding tensor structure in coordinate space takes the form,
\begin{equation}\label{eq:16}
xW^{ij}(x_g, r_\perp) = \frac{1}{2}\delta^{ij}xG(x_g, r_\perp) + \frac{1}{2}\left(\delta^{ij}-2\frac{r^i_\perp r^j_\perp}{r^2_\perp}\right)xH(x_g, r_\perp)
\end{equation}
where $xG(x_g, r_\perp)$ and $xH(x_g, r_\perp)$ are unpolarized and linearly polarized WW gluon distributions in coordinate space.

Combining \eq{eq:15} and \eq{eq:16}, then the cross section becomes
\begin{align}
\frac{\ud\sigma^{\gamma^\ast_TA\to q\bar{q}X}}{\ud \mathcal{P.S.}}=&\alpha_\text{em}\alpha_s e^2_q\delta(x_{\gamma^\ast}-1)z(1-z)\frac{1}{(P^2_\perp+\epsilon^2_f)^4}\nonumber \\
&\times \left\lbrace\left[\left(z^2+(1-z)^2\right)(P^4_\perp+\epsilon^4_f)+2m^2_qP^2_\perp\right]\wt{xG}(x_g,P_\perp,q_\perp)\right. \nonumber \\
&\qquad \left.-\left[2\left(z^2+(1-z)^2\right)P^2_\perp\epsilon^2_f-2m^2_qP^2_\perp\right]\cos(2\Delta\phi)\wt{xH}(x_g,P_\perp,q_\perp)\right\rbrace ,\label{eq:17} \\
\frac{\ud\sigma^{\gamma^\ast_L A\to q\bar{q}X}}{\ud \mathcal{P.S.}}=&8\alpha_\text{em}\alpha_s e^2_q Q^2\delta(x_{\gamma^\ast}-1)z^3(1-z)^3\frac{P^2_\perp}{(P^2_\perp+\epsilon^2_f)^4}\left(\wt{xG}(x_g,P_\perp,q_\perp)+\cos(2\Delta\phi)\wt{xH}(x_g,P_\perp,q_\perp)\right).\label{eq:18}
\end{align}
where $\Delta\phi$ is the angle between $P_\perp$ and $q_\perp$, $\ud\mathcal{P.S.}=\ud^3k_1\ud^3k_2$, and 
\begin{align}
\wt{xG}(x_g, P_\perp, q_\perp) &= \int \frac{\ud^2 r_\perp}{(2\pi)^2}e^{-S_\text{sud}(P_\perp, r_\perp)}e^{-iq_\perp\cdot r_\perp}xG(x_g, r_\perp) \label{eq:19} \\
\wt{xH}(x_g, P_\perp, q_\perp) &= \int \frac{\ud^2 r_\perp}{(2\pi)^2}e^{-S_\text{sud}(P_\perp, r_\perp)}e^{-iq_\perp\cdot r_\perp}xH(x_g, r_\perp)
\end{align}
$\wt{xG}(x_g, P_\perp, q_\perp)$ and $\wt{xH}(x_g, P_\perp, q_\perp)$ are the resumed WW gluon distributions in momentum space. From the above expressions, one can clearly see that the two distributions have resum small-$x$ logarithms and Sudakov double logarithms simultaneously. We would like to point out that the resummations performed here is similar to that in Refs.~\cite{Mueller:2012uf,Mueller:2013wwa}, alternatively, a more general description of TMD evolutions has been done in \cite{Xiao:2017yya}, in which the two resummations can be performed without regarding in a specific scattering process in the CGC formalism. Notice that, for the linearly polarized WW gluon distribution, $\wt{xH}(x_g, P_\perp, q_\perp)$, one can not perform this transformation directly, this is mainly due to the tensor structure $\frac{1}{2}\left(\delta^{ij}-2\frac{r^i_\perp r^j_\perp}{r^2_\perp}\right)$ is hidden in $xH(x_g, r_\perp)$. In practice, when performing this transformation, a factor of $-\cos(2\theta)$ will be generated, where $\theta$ is the angle between $q_\perp$ and $r_\perp$.

From \eq{eq:17} and \eq{eq:18}, one can see that the differential cross section can be formulated in terms of $\wt{xG}$ and $\wt{xH}$. Especially, the coefficients of $\cos(2\Delta\phi)$ terms in the above expressions can provide us the direct information of linearly polarized WW gluon distributions, $\wt{xH}$. If one integrate out $\phi$, then the coefficients will disappear and the cross section only dependent on the unpolarized ones, $\wt{xG}$. This indicates that the two gluon distributions can be extracted by measuring the total cross section and the coefficient of $\cos(2\Delta\phi)$ respectively. As argued in Refs.~\cite{Metz:2011wb,Dominguez:2011br}, since we are working in coordinate space, $\Delta\phi=\phi_u-\phi_v$ with $\phi_u$ and $\phi_v$ being the azimuthal angles of $\bo{u}$ and $\bo{v}$, while $P_\perp$ and $q_\perp$ are the conjugate momenta of $\bo{u}$ and $\bo{v}$, therefore, it is transparent to explore the properties of $\wt{xH}$ by measuring the asymmetry of $P_\perp$ and $q_\perp$. Furthermore, attention should be paid is that, for transversely and longitudinally polarized photons, there exists a sign change for the coefficients of $\cos(2\Delta\phi)$, as shown in \eq{eq:17} and \eq{eq:18}. Especially, if $Q\approx 0$ (for quasi-photon scattering process), the coefficient will become a negative one.

\section{General properties of WW gluon distributions in McLerran-Venugopalan model}\label{sec:ww}
%==============================================================================
In the last section, we have discussed the quark-antiquark pair (or dijet) cross section that can be formulated in terms of $\wt{xG}$ and $\wt{xH}$, and pointed out that the $\cos(2\Delta\phi)$ coefficients directly relate to the linearly polarized ones. In order to calculate the cross sections, the explicit expressions of $\wt{xG}$ and $\wt{xH}$ should be specified. In what follows, we do not include the Sudakov factor and will apply the original McLerran-Venugopalan (MV) model~\cite{McLerran:1993ka,McLerran:1993ni,McLerran:1994vd} to study the behaviors of the two gluon distributions. In this model, the color charge density $\rho_a(x^+,\bo{x})$ conforms to a Gaussian distribution and satisfies the following relation,
\begin{equation}\label{eq:21}
\left\langle\rho_a(x^+,\bo{x})\rho_b(y^+,\bo{y})\right\rangle_{x_g}=\delta_{ab}\delta^{(2)}(\bo{x}-\bo{y})\mu^2_{x_g}(x^+),
\end{equation}
where $\mu^2_{x_g}$ is the squared density of color charges per area. The above relation only hold when quantum evolution effects are not large, but $x_g$ is sufficiently small such that the high density regime can be accessed. In this sense, the color charge density $\rho_a(x^+,\bo{x})$ is regarded as static source, while for small-$x$ gluons, they are treated as classical gauge field $A^-_a$ due to the density is very large and can be expressed in terms of $g_s\rho$ via the Yang-Mills equation,
\begin{equation}
A^-_a(x^+,\bo{x})=g_s\int \ud^2\bo{z} G_0(\bo{x}-\bo{z})\rho_a(x^+,\bo{z}), \quad G_0(\bo{x})=\int \frac{\ud^2k_\perp}{(2\pi)^2}\frac{e^{ik_\perp\cdot\bo{x}}}{k^2_\perp}
\end{equation}
where $G_0(\bo{x})$ is the gluon propagator. Using \eq{eq:21}, one can obtain the two-point function of gluon field $A^-_a$,
\begin{equation}
\left\langle A^-_a(x^+,\bo{x})A^-_b(y^+,\bo{y})\right\rangle_{x_g}=\frac{1}{g^2_s}\delta_{ab}\delta(x^+-y^+)\mu^2_{x_g} L_{\bo{x}\bo{y}},
\end{equation}
with
\begin{equation}
L_{\bo{x}\bo{y}}=g^4_s\int \ud^2 \bo{z}G_0(\bo{x}-\bo{z})G_0(\bo{y}-\bo{z})=g^4_s\int \frac{\ud^2 l_\perp}{(2\pi)^2}\frac{1}{l^4_\perp}e^{il_\perp\cdot(\bo{x}-\bo{y})}.
\end{equation}
By introducing a dimensional function
\begin{equation}
\Gamma(\bo{x}-\bo{y})\equiv \mu^2_{x_g}(L_{\bo{x}\bo{x}} + L_{\bo{y}\bo{y}} - 2L_{\bo{x}\bo{y}}).
\end{equation}
It is straightforward to show that the quark dipole reads,
\begin{equation}
S^{(2)}_{q\bar{q}}(\bo{x},\bo{y})\equiv\frac{1}{N_c}\left\langle\text{Tr}U(\bo{x})U^\dagger(\bo{y})\right\rangle_{x_g}=e^{-\frac{C_F}{2}\Gamma(\bo{x}-\bo{y})},
\end{equation}
and the gluon dipole as,
\begin{equation}
S^{(2)}_{gg}(\bo{x},\bo{y})\equiv\frac{1}{N^2_c-1}\left\langle\text{Tr}W(\bo{x})W^\dagger(\bo{y})\right\rangle_{x_g}=e^{-\frac{N_c}{2}\Gamma(\bo{x}-\bo{y})},
\end{equation}
with
\begin{equation}
\begin{split}
\Gamma(\bo{x}-\bo{y}) &= 2g^4_s\mu^2_{x_g}\int\frac{\ud^2 l_\perp}{(2\pi)^2}\frac{1}{l^4_\perp}\left(1-e^{il_\perp\cdot(\bo{x}-\bo{y})}\right) \\
&\approx\alpha_s(g_s\mu_{x_g})^2 (\bo{x}-\bo{y})^2\ln\frac{1}{\vert \bo{x}-\bo{y}\vert\Lambda_{\text{QCD}}}\\
&=\frac{1}{2C_F} (\bo{x}-\bo{y})^2 Q^2_s\ln\frac{1}{\vert \bo{x}-\bo{y}\vert\Lambda_{\text{QCD}}}
=\frac{1}{2N_c} (\bo{x}-\bo{y})^2 Q^2_{sg}\ln\frac{1}{\vert \bo{x}-\bo{y}\vert\Lambda_{\text{QCD}}}.
\end{split}
\end{equation}
where $Q^2_s=2C_F\alpha_s g^2_s \mu^2_{x_g}$ and $Q^2_{sg}=2N_c\alpha_s g^2_s\mu^2_{x_g}$ are the quark and gluon saturation scales, respectively. $\Lambda_{\text{QCD}}$ introduced here is the QCD infrared scale. It should be noted that the logarithmic term in the above expressions is very important and is the crucial difference between the Golec-Biernat-W\"usthoff (GBW)~\cite{GolecBiernat:1998js} and the MV model. Particularly, the logarithmic terms in MV model restores the correct high $k_\perp$ perturbative behaviors for the WW gluon distributions, $\sim Q^2_s/k^2_\perp$ (see for examples~\cite{Gelis:2001da,Kharzeev:2003wz}). For the remainder of this work, we will use the functional form
\begin{equation}
\Gamma(\bo{x}-\bo{y})=\frac{1}{2C_F}(\bo{x}-\bo{y})^2 Q^2_s\ln\left(\frac{1}{\vert\bo{x}-\bo{y}\vert\Lambda_{\text{QCD}}}+e\right)
\end{equation}
and choose $\Lambda_{\text{QCD}}=0.241\ \mathrm{GeV}$, $Q^2_{s0}=0.2\ \mathrm{GeV}^2$ as the inputs of rcBK equation. The two values used here are obtained by fitting the structure function $F_2$ with the HERA DIS $ep$ data. We stress that the initial saturation scale $Q^2_{s0}$ is that of proton's at $x_0=0.01$, for heavy nucleus target by using $Q^2_{s0A}=A^{1/3}Q^2_{s0}$. 

Due to the fact that the saturation scale will increase with decreasing $x_g$, in what follows, we use a model independent definition via the relation~\cite{Lappi:2013zma},
\begin{equation}
S^{(2)}_{q\bar{q}}(x_g,\vert\bo{x}-\bo{y}\vert=\sqrt{2}/Q_s)=e^{-1/2}.
\end{equation}
In this sense, $Q^2_s$ can be obtained by using the solution of rcBK equation, while, it is only the saturation scale in the fundamental representation, the gluon saturation scale can be obtained by using $Q^2_{sg}=N_c/C_F Q^2_s$.

Another important ingredient is the double derivative of quadrupole correlator, which is directly relate to the definition of WW gluon distribution, \eq{eq:15}, and can be expressed as~\cite{Dominguez:2011wm,Marquet:2017xwy}
\begin{equation}
\frac{1}{N_c}\left\langle\text{Tr}\left[\partial^iU(\bo{v})\right]U^\dagger(\bo{v}')\left[\partial^jU(\bo{v}')\right]U^\dagger(\bo{v})\right\rangle_{x_g}=\frac{C_F}{N_c}\frac{1-e^{-\frac{N_c}{2}\Gamma(\bo{v}-\bo{v}')}}{\Gamma(\bo{v}-\bo{v}')}\frac{\partial}{\partial\bo{v}^i}\frac{\partial}{\partial\bo{v}'^j}\Gamma(\bo{v}-\bo{v}')
\end{equation}
with
\begin{equation}
\frac{\partial}{\partial\bo{v}^i}\frac{\partial}{\partial\bo{v}'^j}\Gamma(\bo{v}-\bo{v}')=-2g^4_s\mu^2_{x_g}\int\frac{\ud^2l_\perp}{(2\pi)^2}\frac{l^i_\perp l^j_\perp}{l^4_\perp}e^{il_\perp\cdot(\bo{v}-\bo{v}')}
\end{equation}

\subsection{Unpolarized and linearly polarized WW gluon distributions in coordinate space}
%==============================================================================
In order to give a systematical study, we firstly express the two gluon distributions in coordinate space. By collecting \eq{eq:15} and performing the projection onto $\frac{1}{2}\delta^{ij}$ and $\frac{1}{2}(\delta^{ij}-2\frac{r^i_\perp r^j_\perp}{r^2_\perp})$, one can directly obtain,
\begin{equation}\label{eq:unpolarized-coordinate}
\begin{split}
xG(x_g, r_\perp=\bo{v}-\bo{v}') &= -\frac{2S_\perp}{\alpha_s (2\pi)^2}\delta^{ij}\left\langle\text{Tr}\left[\partial^i U(\bo{v})\right]U^\dagger(\bo{v}')\left[\partial^j U(\bo{v}')\right]U^\dagger(\bo{v})\right\rangle_{x_g}\\
&=\frac{2C_F S_\perp}{\alpha_s\pi^2}\frac{1}{(\bo{v}-\bo{v}')^2}\left[1-\exp\left(-\frac{1}{4}(\bo{v}-\bo{v}')^2 Q^2_{sg}\ln\frac{1}{\vert \bo{v}-\bo{v}'\vert \Lambda_{\text{QCD}}}\right)\right]
\end{split}
\end{equation}
and
\begin{equation}\label{eq:polarized-coordinate}
\begin{split}
xH(x_g, r_\perp=\bo{v}-\bo{v}') &= -\frac{2S_\perp}{\alpha_s (2\pi)^2}\left[\delta^{ij}-2\frac{(\bo{v}-\bo{v}')^i (\bo{v}-\bo{v}')^j}{(\bo{v}-\bo{v}')^2}\right]\left\langle\text{Tr}\left[\partial^i U(\bo{v})\right]U^\dagger(\bo{v}')\left[\partial^j U(\bo{v}')\right]U^\dagger(\bo{v})\right\rangle_{x_g}\\
&=\frac{C_F S_\perp}{\alpha_s \pi^2}\frac{1}{(\bo{v}-\bo{v}')^2\ln\frac{1}{\vert\bo{v}-\bo{v}'\vert\Lambda_{\text{QCD}}}} \left[1-\exp\left(-\frac{1}{4}(\bo{v}-\bo{v}')^2 Q^2_{sg}\ln\frac{1}{\vert\bo{v}-\bo{v}'\vert\Lambda_{\text{QCD}}}\right)\right]
\end{split}
\end{equation}
To investigate the general properties of unpolarized and linearly polarized WW gluon distributions in coordinate space, we typically distinguish two regions. For small dipole size limit, $\vert r_\perp\vert \ll 1/Q_{sg}$, one can easily check that $xG(x_g, r_\perp)\simeq \frac{C_F S_\perp Q^2_{sg}}{2\alpha_s \pi^2}\ln\frac{1}{\vert r_\perp\vert\Lambda_{\text{QCD}}}$ and $xH(x_g, r_\perp)\simeq \frac{C_F S_\perp Q^2_{sg}}{4\alpha_s \pi^2}$. In the small $r_\perp$ limit, $xG(x_g, r_\perp)$ behaves as $\ln\frac{1}{\vert r_\perp\vert\Lambda_{\text{QCD}}}$ and $xH(x_g, r_\perp)$ to be a constant at given $x_g$, the perturbative behaviors are actually in agreement with the properties at large transverse momentum~\cite{Sun:2011iw}. Moreover, it is interesting to note that the two gluon distributions have strong dependence on $A$ since $Q^2_{sg}\sim A^{1/3}$ and $S_\perp \sim R^2_A\sim A^{2/3}$, which indicates, in the small $r_\perp$ limit, the two gluon distributions have no nuclear shadowing effect, this feature also holds at large transverse momentum. Furthermore, it should be noted that $\frac{xH(x_g, r_\perp)}{xG(x_g, r_\perp)}\simeq \frac{1}{2}\ln^{-1}\frac{1}{\vert r_\perp\vert\Lambda_{\text{QCD}}}$, the ratio seems to be zero when $r_\perp$ is small enough even though $xH$ still remains non-zero value. Last but not least, the evolutions of the two gluon distributions toward to small-$x$ are completely captured by $Q^2_{sg}$. However, for $\vert  r_\perp\vert\gg 1/Q_{sg}$, but $\vert r_\perp\vert\leq 1/\Lambda_{\text{QCD}}$, one can check that $xG(x_g, r_\perp)\simeq \frac{2C_F S_\perp}{\alpha_s \pi^2}\frac{1}{r^2_\perp}$ and $xH(x_g, r_\perp)\simeq \frac{C_F S_\perp}{\alpha_s \pi^2}\frac{1}{r^2_\perp\ln\frac{1}{\vert r_\perp\vert\Lambda_{\text{QCD}}}}$, the two gluon distributions are independent on $x_g$ and fall off as $r^2_\perp$ with increasing $\vert r_\perp\vert$, these behaviors are completely different from that in small $r_\perp$ limit. Also, the ratio $\frac{xH(x_g, r_\perp)}{xG(x_g, r_\perp)}\approx 1/2$, which indicates the linearly polarized WW gluon distribution will tend to its saturated boundary\footnote{In practice, the saturated boundary is sensitive to the treatment of logarithmic term $\ln\frac{1}{\vert r_\perp\vert\Lambda_{\text{QCD}}}$. If one absorb $\frac{1}{2}$ into the logarithmic term, then the saturated boundary tends to 1.} when $r_\perp$ is large enough. 

In order to give numerical results of unpolarized and linearly polarized WW gluon distributions in coordinate space, according to \eq{eq:unpolarized-coordinate} and \eq{eq:polarized-coordinate}, gluon saturation scale $Q^2_{sg}$ should be known, this value can be obtained by solving the rcBK equation. To avoid repetition, details can be found in Refs.~\cite{Albacete:2010sy,Zhao:2013kea}. We stress that the key input of rcBK equation is the initial saturation scale $Q^2_{s0}$, in what follows, we choose $Q^2_{s0}=0.2\ \mathrm{GeV}^2$ for proton.

\begin{figure}[hbt]
\includegraphics[width=\linewidth]{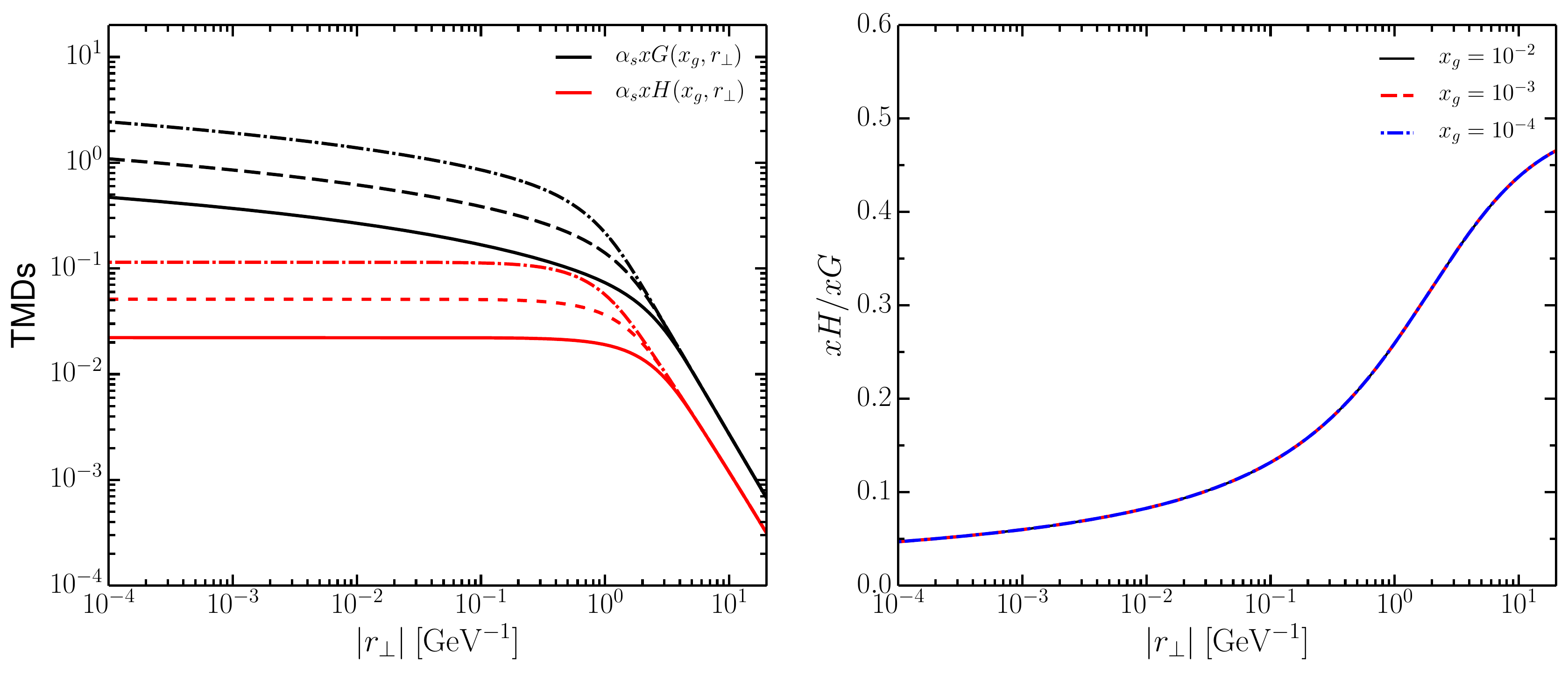}
\caption{Unpolarized and linearly polarized WW gluon distributions in coordinate space. Black and red curves denote the unpolarized and the linearly polarized WW gluon distributions, solid, dashed and dot-dashed lines represent the results at $x_g=10^{-2}, 10^{-3}$ and $10^{-4}$, respectively. Right panel shows the ratio $xH(x_g, r_\perp)/xG(x_g, r_\perp)$ as a function of $\vert r_\perp\vert$ for the three different $x_g$.}
\label{fig:fig1}	
\end{figure}

We firstly show the small-$x$ evolutions of the two gluon distributions in coordinate space. In the left panel of \fig{fig:fig1}, we present the numerical results of unpolarized and linearly polarized WW gluon distributions at different values of $x_g$. Black curves denote the unpolarized WW gluon distributions and the red ones for the linearly polarized distributions. Solid, dashed and dot-dashed lines correspond to $x_g=10^{-2}, 10^{-3}$ and $10^{-4}$, respectively. In the small $r_\perp$ region, from this figure, one can clearly see that the unpolarized WW gluon distributions warmly increase with decreasing $r_\perp$, precisely behaves as $\ln\frac{1}{\vert r_\perp\vert\Lambda_{\text{QCD}}}$. While, for the linearly polarized ones, the results tend to flat over a broad small $r_\perp$ region, which indicates the linearly polarized WW gluon distributions still remains non-zero values even when $r_\perp$ is small enough. Besides that, one can observe that both of the two gluon distributions increasing with decreasing $x_g$, the feature is expected since $xG, xH\sim Q^2_{sg}$ and the gluon saturation scale $Q^2_{sg}$ goes large when $x_g$ becomes small. As in large $r_\perp$ region, e.g. $\vert r_\perp\vert\geq 1\ \mathrm{GeV}^{-1}$, one can observe that the unpolarized (linearly polarized) WW gluon distributions respective to different values of $x_g$ merge into a single curve, this is mainly due to the two gluon distributions are independent on $x_g$ when $r_\perp$ is large enough. Especially, in large $r_\perp$ region, one can see that both the two gluon distributions have same behaviors, precisely equal to $r^{-2}_\perp$.

In the right panel of \fig{fig:fig1}, we present the ratio of $xH(x_g, r_\perp)/xG(x_g, r_\perp)$ as a function of $r_\perp$. From this figure, one can observe that the numerical results for the three different values of $x_g$ merge into a single curve, this is mainly due to the ratio becomes a function of $r_\perp$ by comparing \eq{eq:unpolarized-coordinate} and \eq{eq:polarized-coordinate}. Another feature is that the ratio goes large when increasing $r_\perp$ and tends to saturated boundary $1/2$ when $r_\perp$ is large enough, which indicates $xH(x_g, r_\perp)\leq \frac{1}{2}xG(x_g, r_\perp)$. However, for physical aspect, $xH(x_g, r_\perp)\leq xG(x_g, r_\perp)$, this discrepancy mainly comes from the treatment of the logarithmic term in \eq{eq:polarized-coordinate}, $\ln^{-1}\frac{1}{\vert r_\perp\vert\Lambda_{\text{QCD}}}$. In our calculations, we have $\ln^{-1}\frac{1}{\vert r_\perp\vert\Lambda_{\text{QCD}}} \to \ln^{-1}\left(\frac{1}{\vert r_\perp\vert\Lambda_{\text{QCD}}}+e\right)$, if one use $\ln^{-1}\frac{1}{\vert r_\perp\vert^2\Lambda^2_{\text{QCD}}} \to \ln^{-1}\left(\frac{1}{\vert r_\perp\vert^2\Lambda^2_{\text{QCD}}}+e\right)$, which will give the saturated boundary to be 1. We stress that the different treatments will do not alter the following analysis.

\subsection{Unpolarized and linearly polarized WW gluon distributions in momentum space}
%==============================================================================
Correspondingly, the unpolarized and linearly polarized WW gluon distributions can be obtained by using Fourier transformation, they are expressed as follows:
\begin{equation}
\begin{split}
xG(x_g, k_\perp) &= \int\frac{\ud^2 r_\perp}{(2\pi)^2}e^{-ik_\perp\cdot r_\perp} xG(x_g, r_\perp)\\
&=\frac{2C_F S_\perp}{\alpha_s \pi^2}\int\frac{\ud^2 r_\perp}{(2\pi)^2} e^{-ik_\perp\cdot r_\perp}\frac{1}{r^2_\perp}\left[1-\exp\left(-\frac{1}{4}r^2_\perp Q^2_{sg}\ln\frac{1}{\vert r_\perp\vert\Lambda_{\text{QCD}}}\right)\right],
\end{split}
\end{equation}
and
\begin{equation}
\begin{split}
xH(x_g, k_\perp) &= \int\frac{\ud^2 r_\perp}{(2\pi)^2}e^{-ik_\perp\cdot r_\perp}xH(x_g, r_\perp) \\
&=\frac{C_F S_\perp}{\alpha_s \pi^2}\int\frac{\ud^2 r_\perp}{(2\pi)^2}e^{-ik_\perp\cdot r_\perp}\frac{-\cos(2\theta)}{r^2_\perp\ln\frac{1}{\vert r_\perp\vert\Lambda_{\text{QCD}}}} \left[1-\exp\left(-\frac{1}{4}r^2_\perp Q^2_{sg}\ln\frac{1}{\vert r_\perp\vert\Lambda_{\text{QCD}}}\right)\right]\\
&=\frac{C_F S_\perp}{2\alpha_s \pi^3}\int\ud\vert r_\perp\vert\frac{J_2(\vert k_\perp\vert\vert r_\perp\vert)}{\vert r_\perp\vert\ln\frac{1}{\vert r_\perp\vert\Lambda_{\text{QCD}}}}\left[1-\exp\left(-\frac{1}{4}r^2_\perp Q^2_{sg}\ln\frac{1}{\vert r_\perp\vert\Lambda_{\text{QCD}}}\right)\right].
\end{split}
\end{equation}
It is worthwhile to note that the linearly polarized WW gluon distribution cannot be transformed from coordinate space into momentum space directly, this is because the tensor structure $\frac{1}{2}(\delta^{ij}-2\frac{r^i_\perp r^j_\perp}{r^2_\perp})$ is different from $\delta^{ij}$. After performing this projection, the tensor structure will naturally generate a $-\cos(2\theta)$ factor, where $\theta$ is the relative angle between $r_\perp$ and $k_\perp$. Alternatively, $xG(x_g, k_\perp)$ and $xH(x_g, k_\perp)$ can be arrived at by projecting $xW^{ij}$ onto $\frac{1}{2}\delta^{ij}$ and $\frac{1}{2}(2\frac{k^i_\perp k^j_\perp}{k^2_\perp}-\delta^{ij})$, respectively, which has been done in Refs.~\cite{Metz:2011wb,Dominguez:2011br}.

In the above expression, $J_2(\vert k_\perp\vert \vert r_\perp\vert)$ is the Bessel function of the first kind. The following analysis is similar to that for gluon distributions in coordinate space. For $k^2_\perp \gg Q^2_{sg}$, the dominated contribution comes from the small dipole size $\vert r_\perp\vert \ll 1/Q_{sg}$, and can be evaluated by expanding the exponential term, then $xG(x_g, k_\perp)\simeq \frac{C_F S_\perp}{4\alpha_s \pi^3}\frac{Q^2_{sg}}{k^2_\perp}$ and $xH(x_g, k_\perp)\simeq \frac{C_F S_\perp}{4\alpha_s \pi^3} \frac{Q^2_{sg}}{k^2_\perp}$, in this limit, the two gluon distributions are completely same and behave as $\frac{Q^2_{sg}}{k^2_\perp}$. The perturbative property is agreement with other QCD results~\cite{Gelis:2001da,Kharzeev:2003wz,Nadolsky:2007ba,Catani:2010pd}. Moreover, due to $xG(x_g, k_\perp)\simeq xH(x_g, k_\perp)\simeq\frac{C_F N_c A}{2\pi^3 k^2_\perp}$, the two gluon distributions scale like $A$ and have no nuclear shadowing effects, therefore, each of nucleon can additively give a contribution to the distributions in the dilute region. This conclusion is also agreement with that in coordinate space at small $r_\perp$ limit. In addition, $\frac{xH(x_g, k_\perp)}{xG(x_g, k_\perp)}\approx 1$ when $k_\perp$ is large enough, which gives a saturated boundary for the linearly polarized WW gluon distribution in the dilute regime~\cite{Metz:2011wb}. For $\Lambda^2_{\text{QCD}}\leq k^2_\perp\ll Q^2_{sg}$, the dominant contribution comes from large distances $r_\perp \gg 1/Q_{sg}$, one can neglect the exponential and the logarithmic $r_\perp$ terms, and arrive at $xG(x_g, k_\perp)\simeq \frac{C_F S_\perp}{2\pi^3\alpha_s}\ln\frac{Q^2_{sg}}{k^2_\perp}$ and $xH(x_g, k_\perp)\simeq \frac{N_c C_F A}{2\pi^3}\frac{1}{Q^2_{sg}}\ln\frac{Q^2_{sg}}{k^2_\perp}$, one can check that the two gluon distributions scale like $A^{2/3}$, which means the two gluon distributions have strong nuclear shadowing effects in the dense regime. Another interesting feature is the linearly polarized WW gluon distribution is weakly dependent on $k_\perp$ and approximates to a constant in this region. Furthermore, one can find $\frac{xH(x_g, k_\perp)}{xG(x_g, k_\perp)}$ is suppressed in the small $k_\perp$ limit, in which the multi-gluon scattering effects play a role that constrains the gluons to be polarized.

\begin{figure}[hbt]
\includegraphics[width=\linewidth]{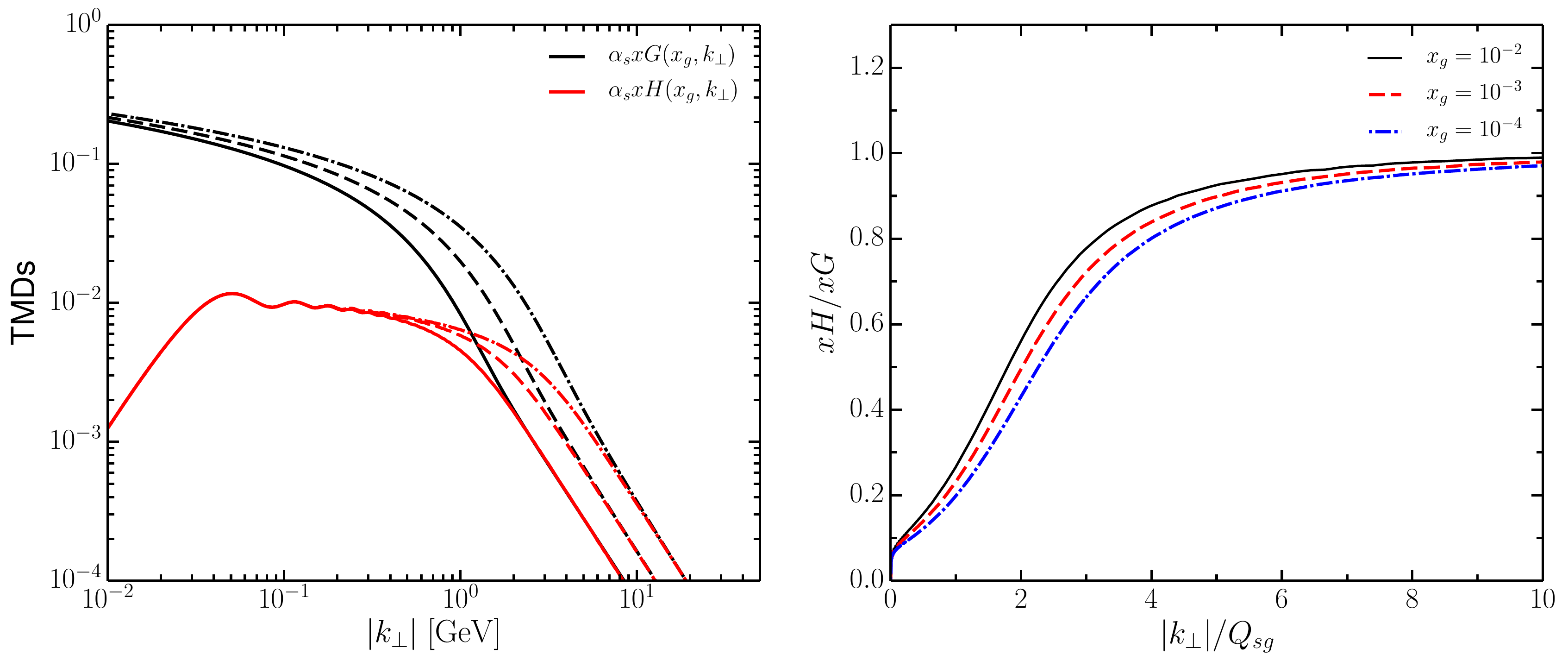}
\caption{Unpolarized and linearly polarized WW gluon distributions in momentum space. Black and red curves denote the unpolarized and the linearly polarized WW gluon distributions, solid, dashed and dot-dashed lines correspond to the results at $x_g=10^{-2}, 10^{-3}$ and $10^{-4}$, respectively. Right panel shows the ratio of $xH(x_g, k_\perp)/xG(x_g, k_\perp)$ as a function of $\vert k_\perp\vert/Q_{sg}$ for the three different $x_g$.}
\label{fig:fig2}	
\end{figure}

In \fig{fig:fig2}, we present the numerical results of the two gluon distributions in momentum space. In the left panel of \fig{fig:fig2}, black and red curves denote the unpolarized and linearly polarized WW gluon distributions, solid, dashed and dot-dashed lines correspond to the results at $x_g=10^{-2}, 10^{-3}$ and $10^{-4}$, respectively. As shown in the left panel of \fig{fig:fig2}, for each value of $x_g$, one can clearly see that the results for unpolarized and linearly polarized WW gluon distributions approach to a single line when $k_\perp$ is very large, which implies the linearly polarized WW gluon distribution tends to its saturated boundary, $xH(x_g, k_\perp)\approx xG(x_g, k_\perp)$. Another important feature is that both the two distributions have same asymptotic behavior, i.e. an inverse power law, precisely equal to $k^{-2}_\perp$ when $k_\perp$ lies in large transverse momentum region. Furthermore, with decreasing $x_g$, one can observe the curves are shifted to large $k_\perp$, this is because, in the large $k_\perp$ region, the evolutions of the two gluon distributions toward to small $x$ are fully characterized by gluon saturation scale, $Q^2_{sg}$, and which is the value becomes large when $x_g$ decreases.  As $k_\perp$ goes small, the two gluon distributions start to separate. Specifically, the unpolarized gluon distributions will warmly increase with decreasing $k_\perp$, while the linearly polarized ones tend to flat over a broad range of $k_\perp$ and are almost independent on $x_g$, this feature depicted here is again confirms what we discussed in the above.

In the right panel of \fig{fig:fig2}, we present the ratio of $xH(x_g, k_\perp)/xG(x_g, k_\perp)$ as a function of $\vert k_\perp\vert/Q_{sg}$ for three different values of $x_g$. The solid, dashed and dot-dashed lines correspond to the results at $x_g=10^{-2}, 10^{-3}$ and $10^{-4}$, respectively. From this figure, one can see that these quantities are weakly dependent on $x_g$ and approx to 1 when $\vert k_\perp\vert /Q_{sg}> 4\div 5$, this strongly indicates the geometric scaling behavior~\cite{Stasto:2000er,Munier:2003vc} is held for the two gluon distributions and the saturated boundary of linearly polarized WW gluon distribution is reached at large $k_\perp$ region. When $k_\perp$ goes small, such as of order of $Q_{sg}$, in which gluon multiple rescattering (saturation) effects to be manifest and play a role that suppressing the gluons to be polarized. Therefore, as shown in this figure, the effects will naturally lead the ratio to be smaller. We note that, although the ratio goes small when decreasing $k_\perp$, it is still sizable even for small $k_\perp$ values, $xH(x_g, k_\perp)/xG(x_g, k_\perp)\geq 10\%$. In this sense, the dijet $\cos(2\Delta\phi)$ structure\footnote{From \eq{eq:17} and \eq{eq:18}, one can check that the $\cos(2\Delta\phi)$ structure is sensitive to the ratio, $xH(x_g, k_\perp)/xG(x_g, k_\perp)$.} in the considered process is expected to be visible even though in small transverse momentum region.

\begin{figure}[hbt]
\includegraphics[width=\linewidth]{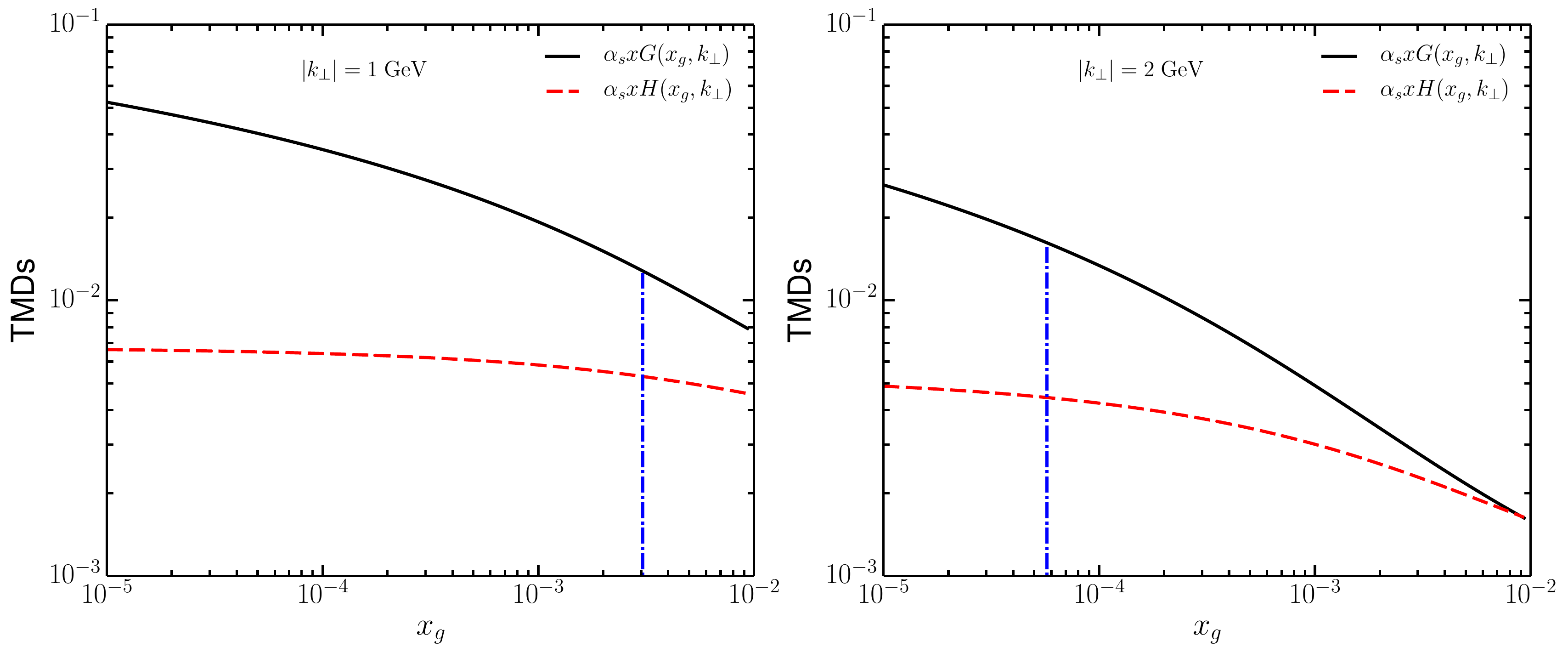}
\caption{Small-$x$ evolutions for unpolarized and linearly polarized WW gluon distributions. Black and red curves denote the unpolarized and linearly polarized ones, left and right panels correspond to the results for $\vert k_\perp\vert=1, 2$ GeV, the vertical dot-dashed lines represent the values of $x_g$ with the condition of $Q_{sg}=\vert k_\perp\vert$, respectively.}
\label{fig:fig3}
\end{figure}

For completeness, we also present small-$x$ evolutions for the two gluon distributions, results are shown in \fig{fig:fig3}. In the calculations, we have chosen two values of $k_\perp$, results presented in the left panel of \fig{fig:fig3} correspond to the case of $\vert k_\perp\vert=1$ GeV, and the right panel is for $\vert k_\perp\vert=2$ GeV. The solid (black) and dashed (red) curves denote unpolarized and linearly polarized WW gluon distributions, and the dot-dashed (blue) vertical lines represent the values of $x_g$ by setting $Q_{sg}=\vert k_\perp\vert$, respectively. In the left panel of \fig{fig:fig3}, one can see that the vertical line separates the domain as two regions, specifically, in which $x_g\geq 0.003$ with $\vert k_\perp\vert\geq Q_{sg}$ as the dilute regime and $x_g\leq 0.003$ with $\vert k_\perp\vert\leq Q_{sg}$ as the dense (saturated) regime. As discussed in the above, we recall that the gluon saturation effects will be very significant in the dense regime ($x_g\leq 0.003$), where gluon rescattering effects will be very important and constrain those gluons to be polarized, therefore, the linearly polarized part becomes flat as shown in the figure. While, in the dilute regime ($x_g\geq 0.003$), the effect is less manifest and makes the linearly polarized part close to the unpolarized one. This feature is more visible for the case of $\vert k_\perp\vert=2$ GeV (see the right panel of \fig{fig:fig3}).

\section{Numerical results and discussions}\label{sec:results}
%==============================================================================
According to \eq{eq:17} and \eq{eq:18}, the quark-antiquark pair (or dijet) cross section is directly related to unpolarized and linearly polarized WW gluon distributions, when the two distributions are known, then the cross section can be obtained. To investigate the azimuthal correlations of dijet in $\gamma^\ast p(A)$ scattering processes, we define the azimuthal correlation function in the following form,
\begin{equation}\label{eq:normal-corr}
C(\Delta\phi)=\frac{\ud\sigma^{\gamma^\ast p \to q\bar{q}X}}{\ud{\mathcal{P.S.}}}\left/\int\ud\Delta\phi\frac{\ud\sigma^{\gamma^\ast p \to q\bar{q}X}}{\ud\mathcal{P.S.}}\right.
\end{equation}
where $\Delta\phi$ is the difference between the azimuthal angles of $P_\perp$ and $q_\perp$. This definition has a clear physical interpretation that quantifying the probability distribution of the dijet versus the relative angle $\Delta\phi$.

To perform numerical calculations, we will focus on typical EIC kinematics and take the center-of-mass energy of the scattering system to be $W=140$ GeV. When the kinematics are specified, $x_g$ enters \eq{eq:17} and \eq{eq:18} and can be determined via,
\begin{equation}\label{eq:xg}
x_g=\frac{1}{W^2+Q^2-M^2_p}\left(Q^2+q^2_\perp+\frac{P^2_\perp}{z(1-z)}\right)
\end{equation}
where $M_p$ is the mass of scattering proton. When performing calculations, we also restrict $\vert P_\perp\vert\geq 2\vert q_\perp\vert$ in order to meet the back-to-back correlation limit. We stress that the chosen kinematics in the following is also relevant for probing gluon saturation effects of scattered target, as we will demonstrate below.

In addition, according to \eq{eq:xg}, $x_g$ increases with increasing $P_\perp$, while it has weak dependence on $q_\perp$, which indicates that $\ud\sigma/\ud\vert P_\perp\vert$ can be used as a proxy to measure the evolutions of $xG(x_g, q_\perp)$ and $xH(x_g, q_\perp)$ towards to $x_g$, while $\ud\sigma/\ud\vert q_\perp\vert$ as a direct probe that measuring the transverse momentum of gluon inside the scattered target.

\subsection{Azimuthal correlations of dijets}
%==============================================================================
\begin{figure}[hbt]
\includegraphics[width=\linewidth]{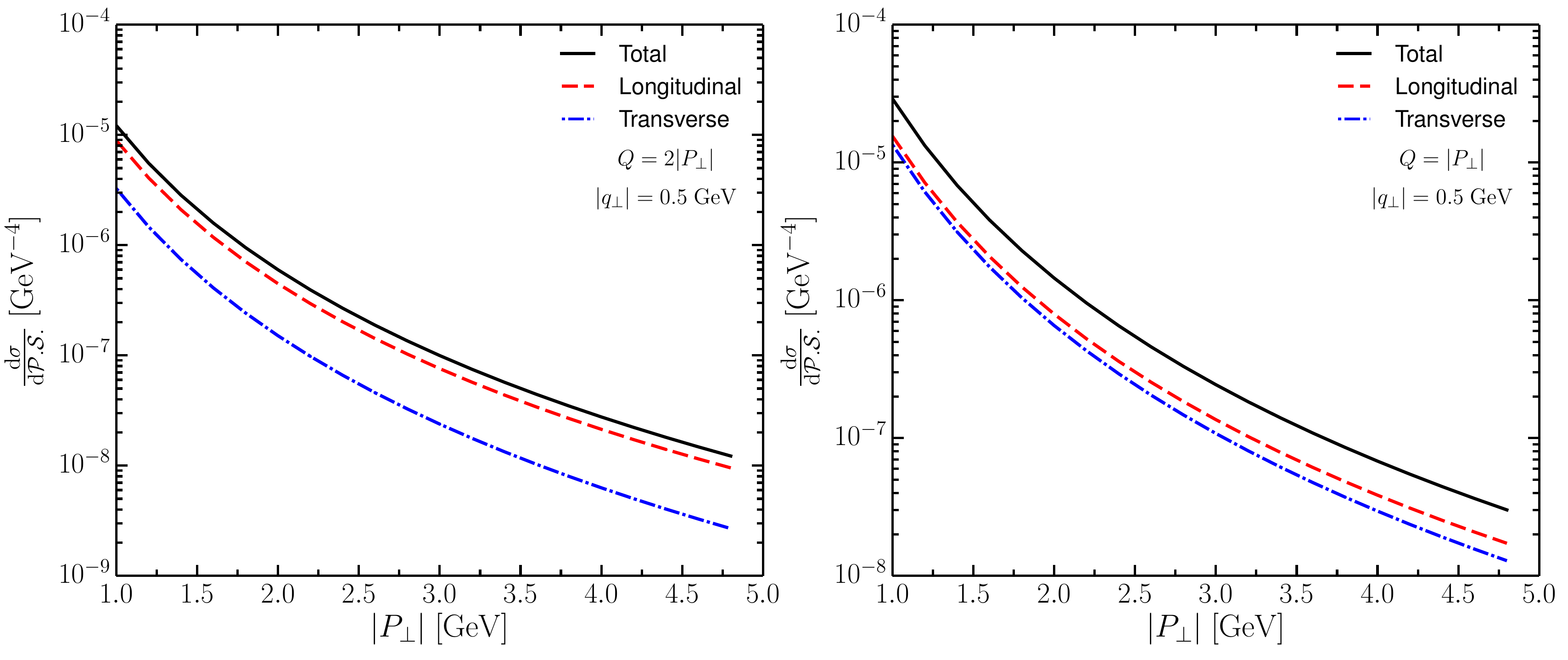}
\caption{Total (black solid), longitudinal (red dashed) and transverse (blue dot-dashed) cross section for light flavored quark dijets as a function of $\vert P_\perp\vert$, where $\vert q_\perp\vert=0.5$ GeV and $z=1/2$. Left panel corresponds to the case of $Q=2\vert P_\perp\vert$ and the right panel is for $Q=\vert P_\perp\vert$.}
\label{fig:fig4}
\end{figure}

In \fig{fig:fig4}, we show the differential cross section for longitudinally and transversely polarized photons as a function of transverse momentum $\vert P_\perp\vert$, in which we have set the transverse momentum imbalance $\vert q_\perp\vert=0.5$ GeV and restrict the dijet to be purely back-to-back, $\Delta\phi=\pi$. For the phenomenological study, we also vary the virtuality of photons as order of transverse momentum, specifically, the left panel in \fig{fig:fig4} shows the results for the case of $Q=2\vert P_\perp\vert$ and the right panel corresponds to the case of $Q=\vert P_\perp\vert$. It also should be reminded that the presented results are for light flavor quark jets with quark mass $m_q=0.01$ GeV and symmetric longitudinal jet $z=1/2$. As can be seen from this figure, the cross sections are rapidly falling with increasing the dijet transverse momentum $\vert P_\perp\vert$. For different virtualities $Q$, the contribution of longitudinally photons and transversely photons to the total cross section is different. When $Q=2\vert P_\perp\vert$ (see left panel of \fig{fig:fig4}), one can observe that the longitudinal part is larger than the transverse part, in this case, the total cross section is mainly driven by the longitudinal part. For another case, when $Q=\vert P_\perp\vert$ (see right panel of \fig{fig:fig4}), although the longitudinal part is still larger than the transverse part, the two parts seem to be comparable in the total cross section. By comparing the two figures, one can conclude that the contribution of transverse part to the total cross section will increase with decreasing the photon's virtualities, $Q^2$.

\begin{figure}[hbt]
\includegraphics[width=\linewidth]{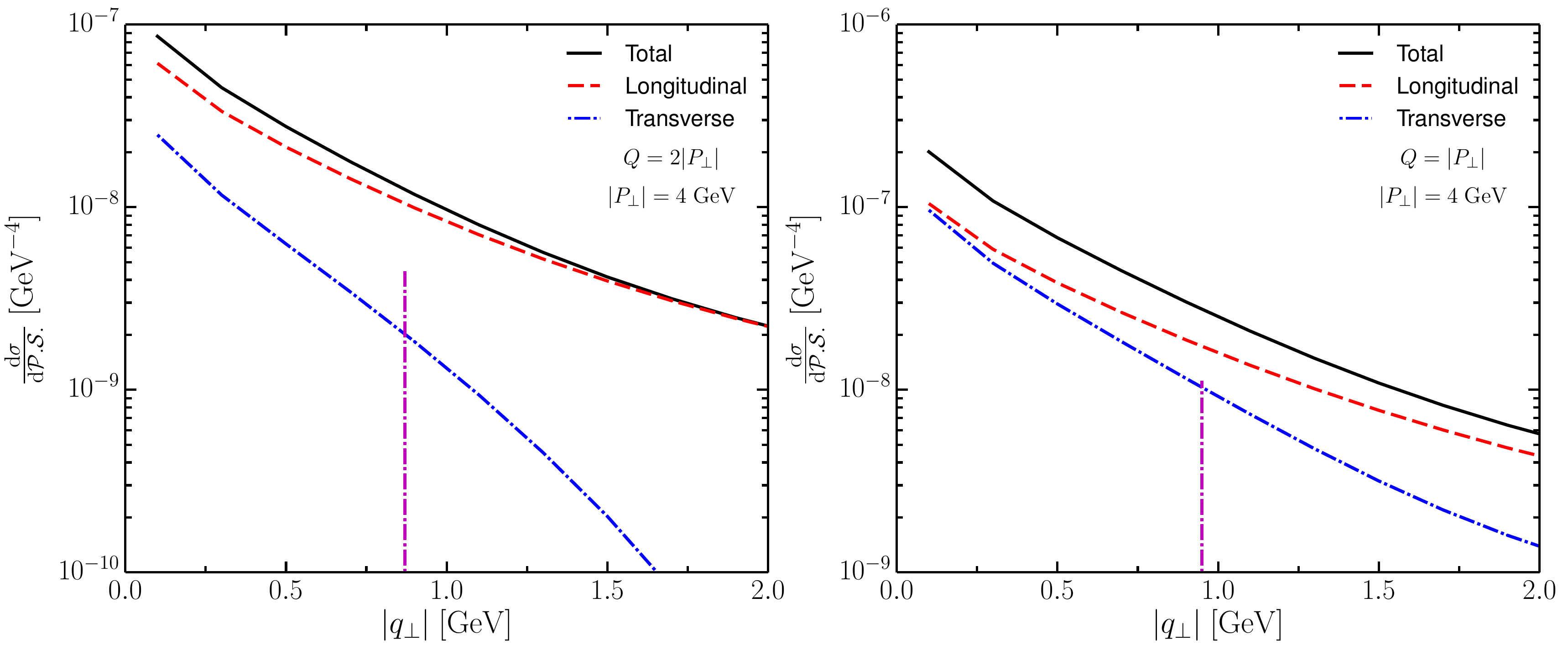}
\caption{Total (black solid), longitudinal (red dashed) and transverse (blue dot-dashed) cross sections for light flavor quark jets as a function of $\vert q_\perp\vert$, where $\vert P_\perp\vert=4$ GeV and $z=1/2$. Left panel corresponds to the case of $Q=2\vert P_\perp\vert$ and the right panel is for $Q=\vert P_\perp\vert$. The vertical lines denote the values of gluon saturation scales at the specific kinematics.}
\label{fig:fig5}
\end{figure}

Similar to \fig{fig:fig4}, in \fig{fig:fig5} we show the differential cross section as a function of transverse momentum imbalance $\vert q_\perp\vert$, in which we have set the transverse momentum $\vert P_\perp\vert=4$ GeV and restrict the dijet to be purely back-to-back, $\Delta\phi=\pi$, the other variables are same as in \fig{fig:fig4}. From this figure, one can observe that the contributions of transversely and longitudinally polarized photons to the total cross section are quite different as comparing to that of in \fig{fig:fig4}. Specifically, for the case of $Q=2\vert P_\perp\vert$ (left panel of \fig{fig:fig5}), one can observe the distribution of transverse part decreases much more sharply than the longitudinal part as increasing $q_\perp$, therefore, the total cross section is almost driven by the longitudinally polarized photons when $q_\perp$ is enough large. The peculiar behaviors exhibited by transversely and longitudinally photons in the figure are not surprising actually, and can be understood as follows: Before to clarify that, it is useful to remind that the transverse momentum imbalance $q_\perp$ is the only variable in the calculations, and as the quantity that directly probes the transverse momentum of incoming gluons from the scattered target. According to \eq{eq:17} and \eq{eq:18}, if only the unpolarized WW gluon distribution $xG(x_g, q_\perp)$ is considered or integrating over $\Delta\phi$, one can expect that the two cross sections are parallelly falling with increasing $q_\perp$. Moreover, it should be noted that the dijets are restricted to be purely back-to-back ($\Delta\phi=\pi$), in this sense, the $\cos(2\Delta\phi)$ terms will give negative (positive) contribution respect for transversely \eq{eq:17} (longitudinally \eq{eq:18}) polarized photons, therefore, the two cross sections will naturally be separated and the separation between the two is directly related to the quantity of $xH(x_g, q_\perp)$ or the ratio of $xH(x_g, q_\perp)/xG(x_g, q_\perp)$. Furthermore, in the specific kinematics, since $x_g$ is weakly dependent on $q_\perp$ and $x_g\approx 0.0065$, correspondingly, the gluon saturation scale $Q_{sg}\approx 0.88$ GeV as shown by the vertical line in the left panel of \fig{fig:fig5}. Reminding that the ratio of $xH(x_g, q_\perp)/xG(x_g, q_\perp)$ becomes large when $q_\perp$ goes large (also see the right panel of \fig{fig:fig2}). As a consequence, the separation between the two cross sections will become more significant as $\vert q_\perp\vert \geq Q_{sg}$, which is shown in the left panel of \fig{fig:fig5}. Similar trends hold for the case of $Q=\vert P_\perp\vert$ (right panel of \fig{fig:fig5}) as well.

As mentioned in previous, the longitudinally polarized photons (\eq{eq:18}) will give a $\cos(2\Delta\phi)$ structure in azimuthal correlations, in contrast, the transversely polarized photons (\eq{eq:17}) gives a $-\cos(2\Delta\phi)$ structure\footnote{If the quark mass is considered, only $Q^2\geq\frac{2m^2_q}{z^2+(1-z)^2}$ will give a negative contribution.}, and this part plays a role that partially washing out the $\cos(2\Delta\phi)$ structure in azimuthal correlations. One can expect that the $\cos(2\Delta\phi)$ structure will become more visible only when the total cross section is dominated by longitudinally polarized photons, otherwise, if the total cross section is mainly driven by the transverse part, it is a challenge to observe the $\cos(2\Delta\phi)$ structure.

\begin{figure}[hbt]
\includegraphics[width=\linewidth]{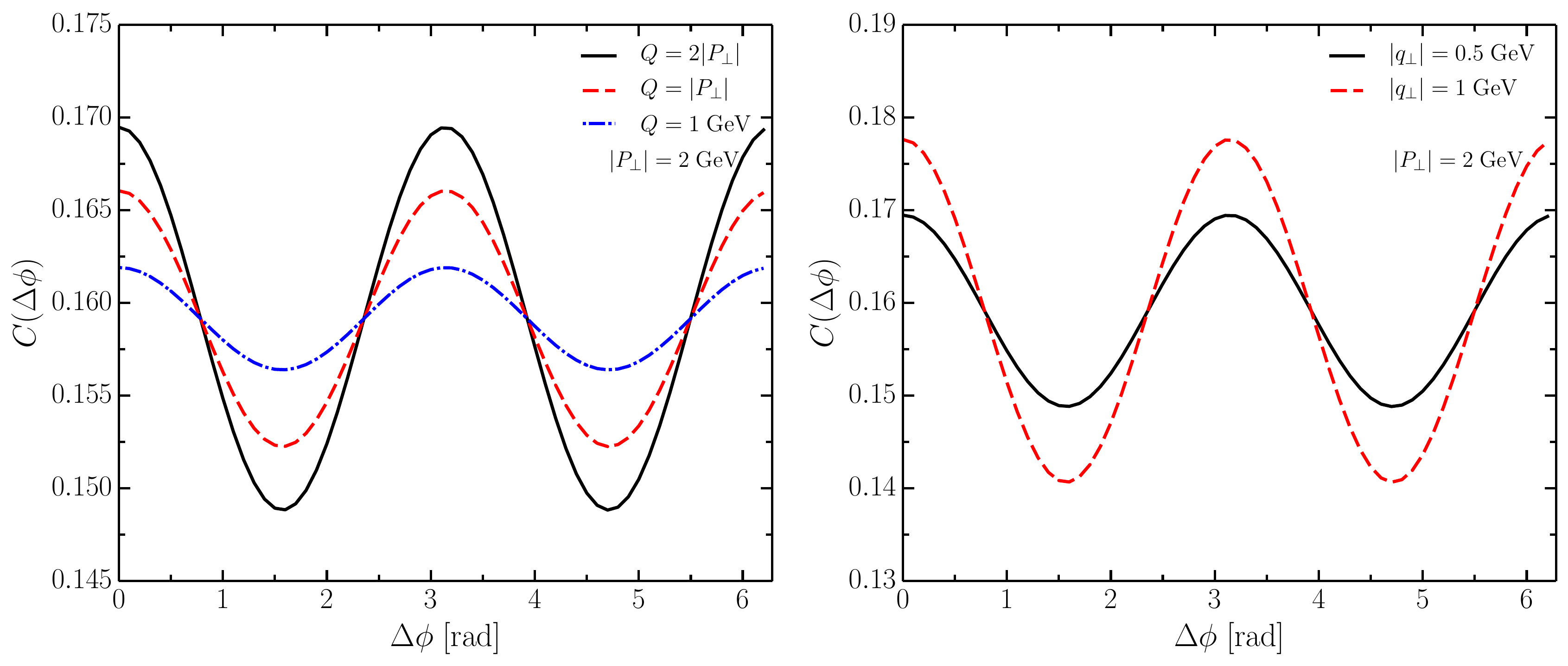}
\caption{The normalized azimuthal correlations $C(\Delta\phi)$ as a function of $\Delta\phi$, in which $W_{\gamma^\ast p}=140$ GeV, $\vert P_\perp\vert=2$ GeV and $z=1/2$. Results in left panel correspond to the case of different photon virtualities $Q$ and the right panel is for different $q_\perp$.}
\label{fig:fig6}
\end{figure}

In \fig{fig:fig6}, we plot the normalized azimuthal correlations $C(\Delta\phi)$ defined in \eq{eq:normal-corr} as a function of relative angle $\Delta\phi$. In the left panel of \fig{fig:fig6}, the represented results are for different photon virtualities $Q$ by setting $\vert P_\perp\vert=2$ GeV and $\vert q_\perp\vert=0.5$ GeV, the other kinematics are same as in \fig{fig:fig4}. The solid (black), dashed (red) and dot-dashed (blue) lines correspond to the case of $Q=2\vert P_\perp\vert, \vert P_\perp\vert$ and 1 GeV, respectively. As can be seen from this figure, the $\cos(2\Delta\phi)$ structure in azimuthal correlations will be suppressed and broaden with decreasing the virtualities $Q$, this is mainly due to the fact that the contribution of longitudinally polarized photons to the total cross section will decrease as $Q$ decreases. In the right panel of \fig{fig:fig6}, we show the azimuthal correlations for different $q_\perp$, in the calculations, we set $Q=2\vert P_\perp\vert$ with $\vert P_\perp\vert=2$ GeV, the other kinematics are same as former. The solid (black) and dashed (red) lines correspond to $\vert q_\perp\vert=0.5$ and 1 GeV, respectively. As discussed in previous, the $\cos(2\Delta\phi)$ structure in azimuthal correlation is sensitive to the ratio of $xH(x_g, q_\perp)/xG(x_g, q_\perp)$, and the ratio will become large as $q_\perp$ goes large, therefore, one can observe that the $\cos(2\Delta\phi)$ structure will become more visible for $\vert q_\perp\vert=1$ GeV as compared to that of $\vert q_\perp\vert=0.5$ GeV.

Furthermore, we have also checked that for the heavy flavor jets, where the heavy quark mass takes $m_q=1.4$ GeV, results are very similar to the light quark jets. For this reason, we do not include separate plots.

\subsection{Sudakov factor will suppress the azimuthal correlations}
%==============================================================================
\begin{figure}[hbt]
\includegraphics[width=\linewidth]{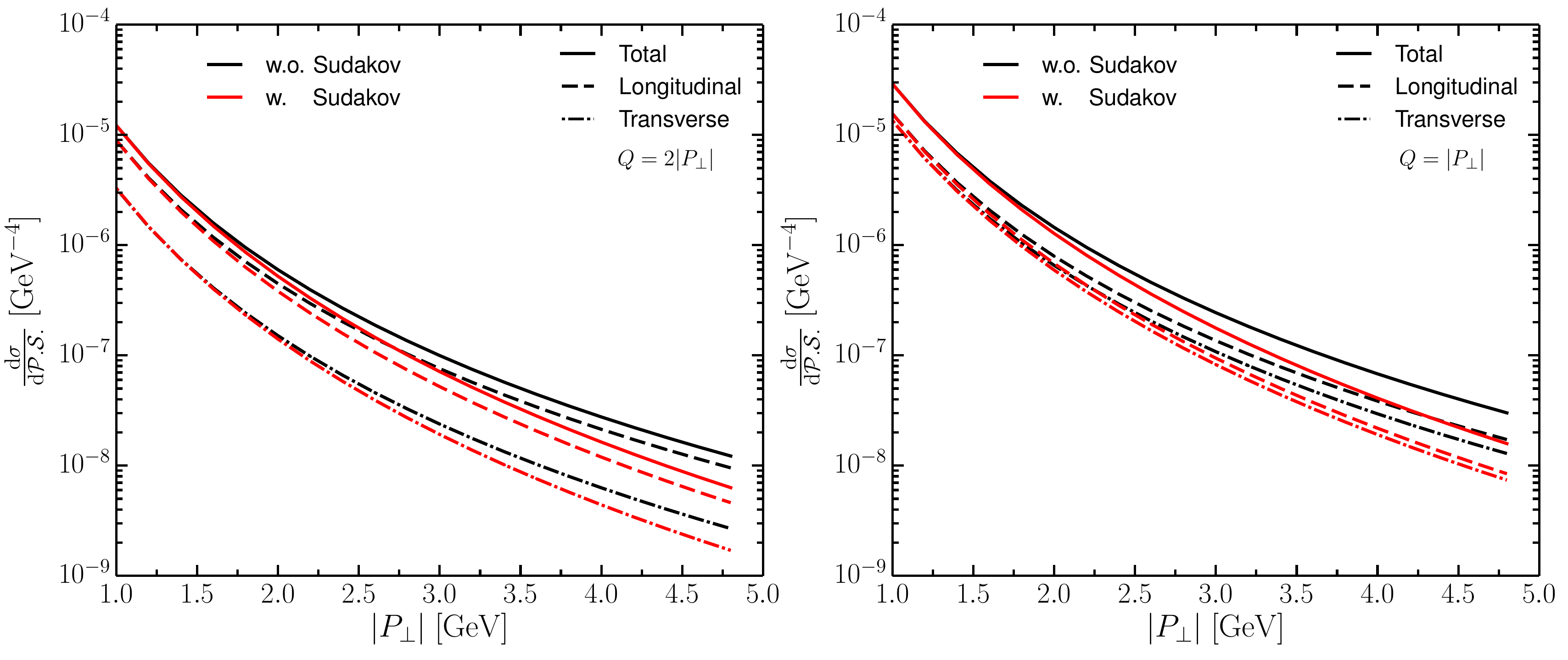}
\caption{Total (solid), longitudinal (dashed) and transverse (dot-dashed) cross sections for dijets as a function of $\vert P_\perp\vert$, where $\vert q_\perp\vert=0.5$ GeV and $z=1/2$. The black and red curves denote without and with Sudakov resummations, left panel corresponds to the case of $Q=2\vert P_\perp\vert$ and the right panel is for $Q=\vert P_\perp\vert$.}
\label{fig:fig7}
\end{figure}

In order to further investigate how does the Sudakov resummations impact on the cross sections, we perform calculations by regarding the same kinematics as in \fig{fig:fig4}, results are shown in \fig{fig:fig7}. Similarly, the left and right panels in \fig{fig:fig7} correspond to the case of $Q=2\vert P_\perp\vert$ and $Q=\vert P_\perp\vert$, the black and red curves denote without and with Sudakov resummations, respectively. By comparing the black and red curves depicted in \fig{fig:fig7}, one can observe that the behavior of the cross sections is not changed after including the Sudakov resummations, the cross sections are tamely falling with increasing $\vert P_\perp\vert$. However, a difference between the two (black and red curves) is that the Sudakov resummations will suppress the cross sections and the suppression tends to be significant when $\vert P_\perp\vert$ goes large, as revealed by the magnitude in the figure. It naturally reflects the fact that the parton shower effect is relatively weaker in the low $\vert P_\perp\vert$ region while the effect goes stronger in the large $\vert P_\perp\vert$ region. This argument is consistent with other perturbative calculations, see e.g. Ref.~\cite{Stasto:2018rci}. Another interesting feature by comparing the black and red curves, such as for the case of $Q=2\vert P_\perp\vert$ (left panel of \fig{fig:fig7}), is that the relative difference between transverse and longitudinal parts goes smaller when the Sudakov resummations are taken into account. More precisely, for the specific kinematics, the ratio of transverse component in the total cross section $\sigma_T/\sigma_{\text{Total}}\approx 0.32$ when including the Sudakov resummations, while $\sigma_T/\sigma_{\text{Total}}\approx 0.25$ if the Sudakov resummations are not included. As discussed in previous, transversely polarized photons gives a contribution that partially washing out the $\cos(2\Delta\phi)$ structure in azimuthal correlations, therefore, one can straightforwardly expect that the $\cos(2\Delta\phi)$ structure will be suppressed and broaden if the Sudakov resummations are taken into account. We further notice that, after including Sudakov resummations, the transverse part seems to be parallelly falling with increasing $P_\perp$ respect to the longitudinal part, in this sense, the $\cos(2\Delta\phi)$ structure in azimuthal correlations will not be changed for different values of $P_\perp$. Similar features hold for the case of $Q=\vert P_\perp\vert$ (see right panel of \fig{fig:fig7}).

\begin{figure}[hbt]
\includegraphics[width=\linewidth]{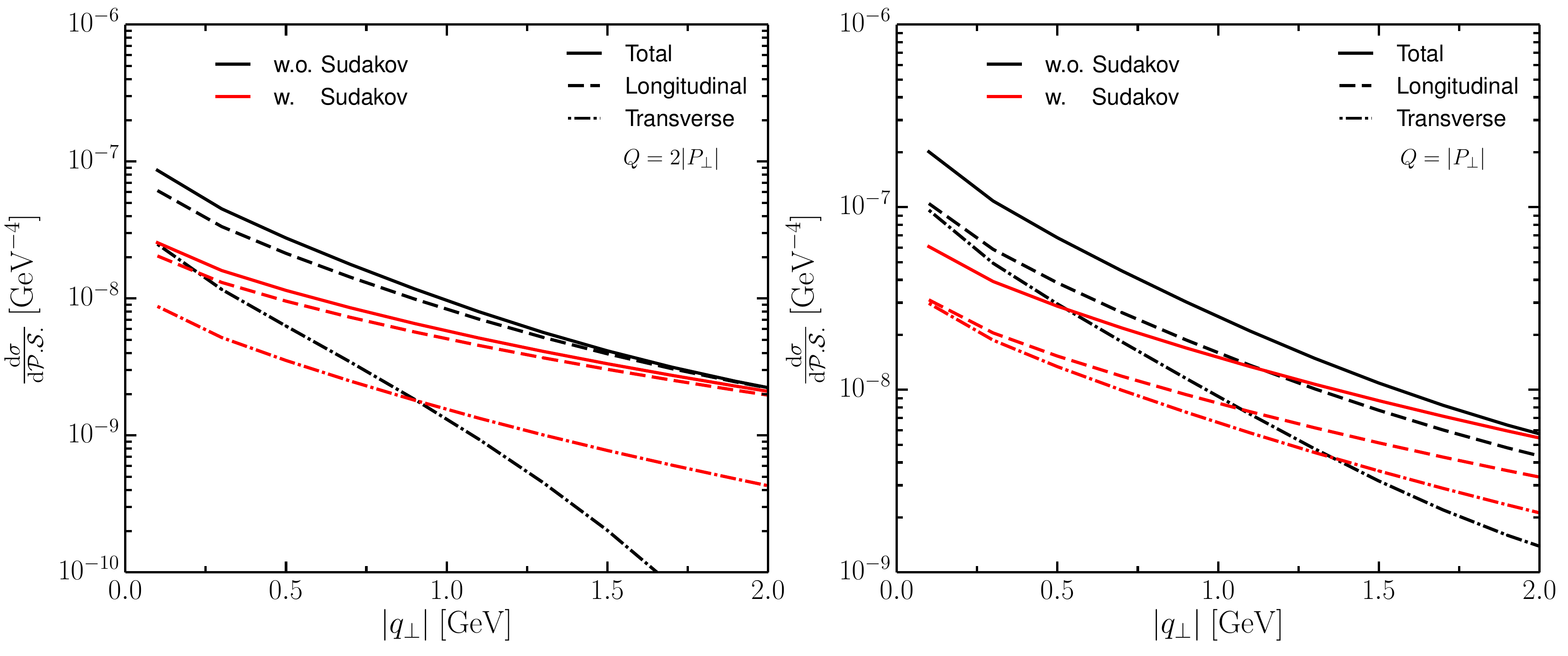}
\caption{Total (solid), longitudinal (dashed) and transverse (dot-dashed) cross sections for the dijet as a function of $\vert q_\perp\vert$, where $\vert P_\perp\vert=4$ GeV and $z=1/2$. The black and red curves denote without and with Sudakov resummations, left panel corresponds to the case of $Q=2\vert P_\perp\vert$ and the right panel is for $Q=\vert P_\perp\vert$.}
\label{fig:fig8}
\end{figure}

In \fig{fig:fig8}, we show the cross sections as a function of transverse momentum imbalance $\vert q_\perp\vert$, the other kinematics are same as in \fig{fig:fig5}. Similarly, the black and red curves denote without and with Sudakov resummations, the left and right panels correspond to the case of $Q=2\vert P_\perp\vert$ and $Q=\vert P_\perp\vert$, respectively. As shown in this figure, the red curves are significantly suppressed as compared to the black one in small $q_\perp$ region, while with increasing $q_\perp$, one can observe this suppression goes small\footnote{For the transverse part, although the distribution is suppressed in small $q_\perp$, one can observe that the distribution with Sudakov resummations is still larger than that of without Sudakov resummations. One should not be confused that, this is because we have set the dijet to be purely back-to-back in the calculations, and the result for the transverse part only quantify the difference between $xG$ and $xH$ (see \eq{eq:17}).}. This actually reflects the fact that Sudakov effects will reduce the distribution at small $q_\perp$ region and make the distributions to be flattened over a broad $q_\perp$ region. We notice that the distributions presented here are somewhat consistent with a recent study in Ref.~\cite{Hatta:2019ixj}, in which the authors have pointed out the Sudakov resummations are very important in describing the dijets transverse momentum imbalance distributions at E791~\cite{Aitala:2000hb,Aitala:2000hc}, and shown only the Sudakov effects is considered that can make the theoretical results meet to the experimental data. Another important feature as shown in the figure is that the difference between transverse and longitudinal cross sections will become smaller if the Sudakov resummations are considered, therefore, the $\cos(2\Delta\phi)$ structure in azimuthal correlations is expected to be suppressed and broaden.

\begin{figure}[hbt]
\includegraphics[width=\linewidth]{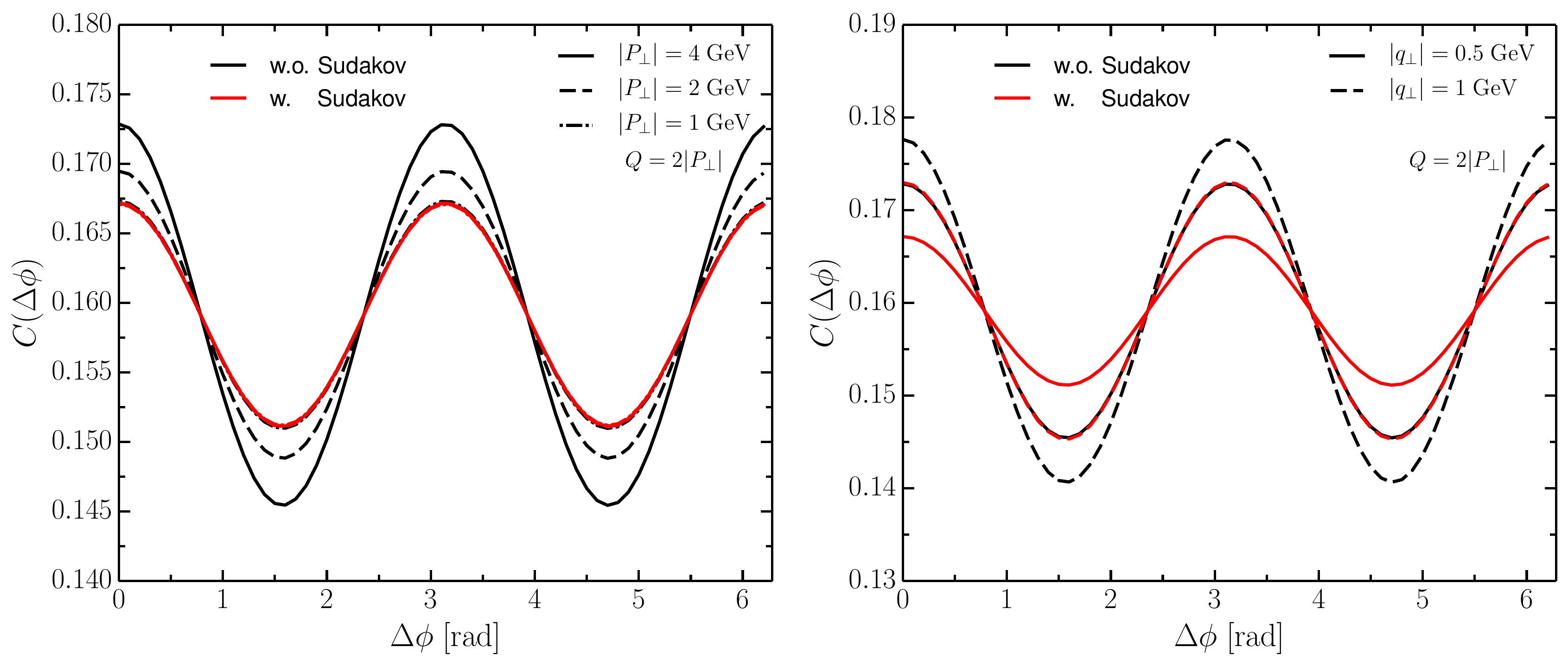}
\caption{The normalized azimuthal correlations $C(\Delta\phi)$ as a function of $\Delta\phi$. Left panel is for difference values of $P_\perp$, and the right panel is for different $q_\perp$.}
\label{fig:fig9}
\end{figure}

In \fig{fig:fig9}, we show the normalized azimuthal correlations $C(\Delta\phi)$ as a function of $\Delta\phi$. In the left panel of \fig{fig:fig9}, we present the results for different $P_\perp$, in which we have set $Q=2\vert P_\perp\vert$, $\vert q_\perp\vert=0.5$ GeV and $z=1/2$. The solid, dashed and dot-dashed lines correspond to $\vert P_\perp\vert=4, 2$ and 1 GeV, the red and black curves denote with and without Sudakov resummations, respectively. From this figure, one can observe that the back-to-back ($\Delta\phi=\pi$) correlations are indeed suppressed when including Sudakov resummations, especially, this suppression goes stronger when $P_\perp$ goes larger. Moreover, in the specific kinematics, one can observe that the red curves are completely overlap together, which indicates the $\cos(2\Delta\phi)$ structure in azimuthal correlations will be unchanged after including Sudakov resummations. This feature can be tracked back to that the transverse cross sections will be parallelly falling respect to the longitudinal ones as $P_\perp$ goes large (see for example \fig{fig:fig7}).

As mentioned in previous, the effect of Sudakov resummations is strongly sensitive to the difference between $\vert P_\perp\vert$ and $\vert q_\perp\vert$, when $Q$ and $\vert P_\perp\vert$ are fixed, one may expect that the effect will decrease with increasing $\vert q_\perp\vert$. In order to verify this, we present results in the right panel of \fig{fig:fig9}, where $Q=4$ GeV and $\vert P_\perp\vert=2$ GeV, the solid and dashed curves correspond to $\vert q_\perp\vert=0.5$ and 1 GeV, the black and red curves denote without and with Sudakov resummations, respectively. From this figure, besides the fact that the Sudakov resummations suppressing the back-to-back correlations, one can observe that the suppression for $\vert q_\perp\vert=0.5$ GeV is stronger than that of $\vert q_\perp\vert=1$ GeV (By comparing the differences between the away-side peaks for the two cases.). Furthermore, it is worthwhile to note that, after including Sudakov resummations, the away-side ($\Delta\phi=\pi$) peak for $\vert q_\perp\vert=1$ GeV is still larger than that of $\vert q_\perp\vert=0.5$ GeV, the main reason comes from the fact that the $\cos(2\Delta\phi)$ structure is sensitive to $xH(x_g, q_\perp)/xG(x_g, q_\perp)$ and the ratio becomes large as $q_\perp$ goes large.

\subsection{Gluon saturation further suppresses the azimuthal correlations of the dijets}
%==============================================================================
As mentioned in previous, the $\cos(2\Delta\phi)$ terms in cross sections directly quantify the values of linearly polarized WW gluon distributions (\eq{eq:17} and \eq{eq:18}), this distribution is strongly suppressed when the gluon saturation dominated.

To further demonstrate the gluon saturation effects will actually suppress the back-to-back correlations, we plan to perform a calculation to compare the dijets azimuthal correlations in $\gamma^\ast p$ and $\gamma^\ast A$ scattering systems with the same kinematics, in which the only difference is the initial saturation scale, for $\gamma^\ast p$ scattering process, the initial saturation scale for proton $Q^2_{s0p}=0.2\ \mathrm{GeV}^2$, for $\gamma^\ast A$, we choose $Q^2_{s0A}=1.2\ \mathrm{GeV}^2$ due to $Q^2_{sA}\sim A^{1/3}Q^2_{sp}$ with $A\approx 200$. Within this assumption, one can expect that the back-to-back correlation in $\gamma^\ast A$ scattering process will be suppressed as compared to that in $\gamma^\ast p$ if gluon saturation effects is the actual reason that smearing the correlations.

\begin{figure}[hbt]
\includegraphics[width=\linewidth]{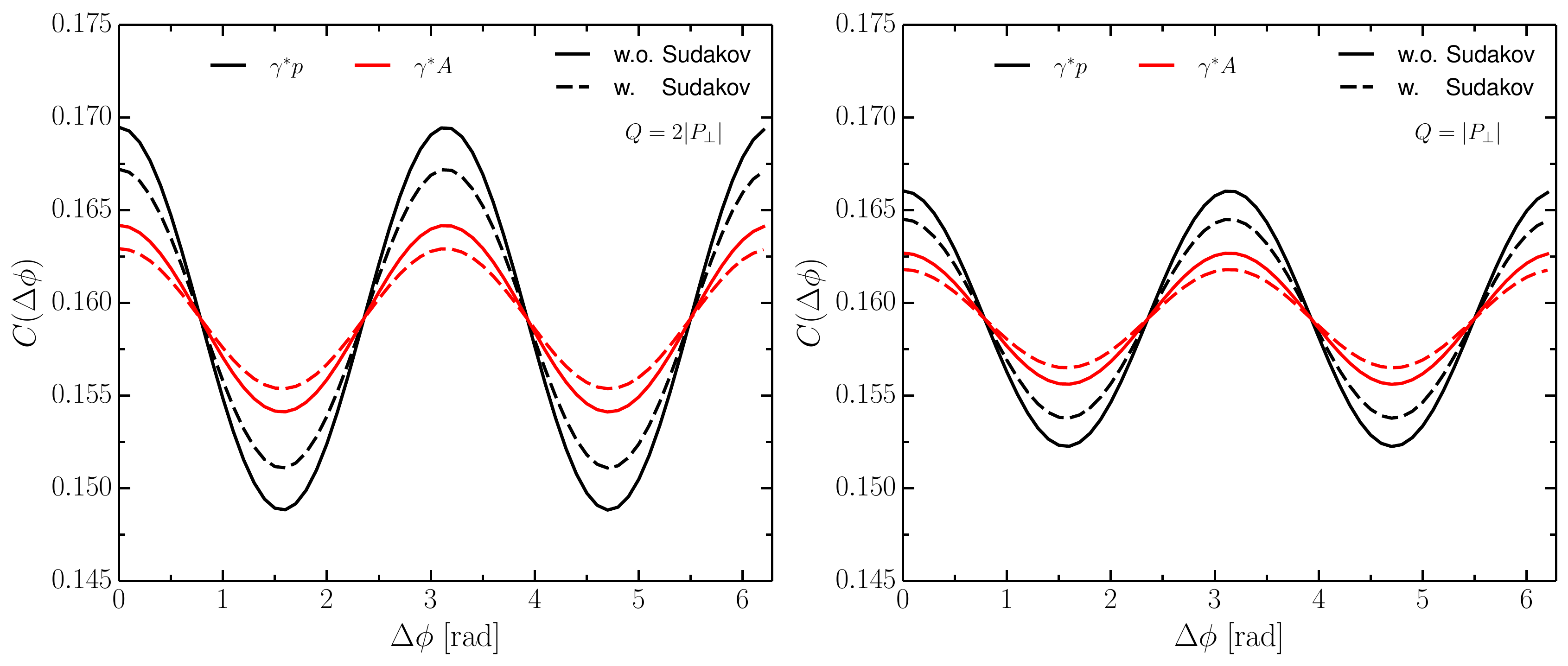}
\caption{The normalized azimuthal correlations function $C(\Delta\phi)$ for different scattering systems, in which $\vert P_\perp\vert=2$ GeV, $\vert q_\perp\vert=0.5$ GeV and $z=1/2$. The black and red curves correspond to $\gamma^\ast p$ and $\gamma^\ast A$, the solid and dashed curves denote without and with Sudakov resummations, respectively. The left panel corresponds to the case of $Q=2\vert P_\perp\vert$ and the right panel is for $Q=\vert P_\perp\vert$.}
\label{fig:fig10}
\end{figure}

In \fig{fig:fig10}, we present the normalized azimuthal correlations for different scattering systems. The black and red curves denote $\gamma^\ast p$ and $\gamma^\ast A$ scattering processes, the solid and dashed lines correspond to without and with Sudakov resummations. The left panel is for $Q=2\vert P_\perp\vert$ and the right panel is for $Q=\vert P_\perp\vert$, in which we have set $\vert P_\perp\vert=2$ GeV, $\vert q_\perp\vert=0.5$ and $z=1/2$. From this figure, by comparing the solid lines with dashed lines, one observe that the back-to-back ($\Delta\phi=\pi$) correlations are suppressed for each colliding system, as discussed in previous, this suppression is mainly come from Sudakov resummations. It should be noted that the chosen kinematics for $\gamma^\ast p$ and $\gamma^\ast A$ systems are same in the calculations, therefore, the effects of Sudakov resummations should be same in the two scattering systems. Furthermore, by comparing the red curves with black ones, one can see that the dijets back-to-back ($\Delta\phi=\pi$) correlations in $\gamma^\ast A$ are significantly suppressed than that of $\gamma^\ast p$. This suppression is mainly due to the gluon saturation effects for heavy ion $A$ is more predominant than the proton $p$. To demonstrate that, it is useful to remind that the enhancement of back-to-back correlations is proportional to the ratio of $\vert q_\perp\vert/Q_{sg}$, and with the fact of gluon saturation scale of $A$ is larger than that of proton ($Q^2_{sA}=A^{1/3}Q^2_{sp}$), therefore, when the kinematics are fixed, the values of $\vert q_\perp\vert/Q_{sg}$ in $\gamma^\ast A$ scattering system is smaller than that in $\gamma^\ast p$, as a consequence, the back-to-back correlations in $\gamma^\ast A$ is naturally suppressed as compared to that in $\gamma^\ast p$ scattering process.

In addition, we also investigate the $\cos(2\Delta\phi)$ anisotropy in the two scattering systems, which is the quantity that can be directly measured in experiments and is defined as follows,
\begin{equation}
v_2\equiv \frac{\int \ud\Delta\phi\cos(2\Delta\phi)\ud \sigma}{\int\ud\Delta\phi\ud \sigma}
\end{equation}
Clearly, the numerator $\int \ud\Delta\phi\cos(2\Delta\phi)\ud\sigma \sim xH(x_g, q_\perp)$, while the denominator $\int\ud\Delta\phi\ud \sigma\sim xG(x_g, q_\perp)$, hence, $v_2\sim xH(x_g, q_\perp)/xG(x_g, q_\perp)$. Therefore, by measuring the elliptic anisotropy $v_2$ of the dijets, we can get the information about $xH/xG$ in the scattered target.

\begin{figure}[hbt]
\includegraphics[width=\linewidth]{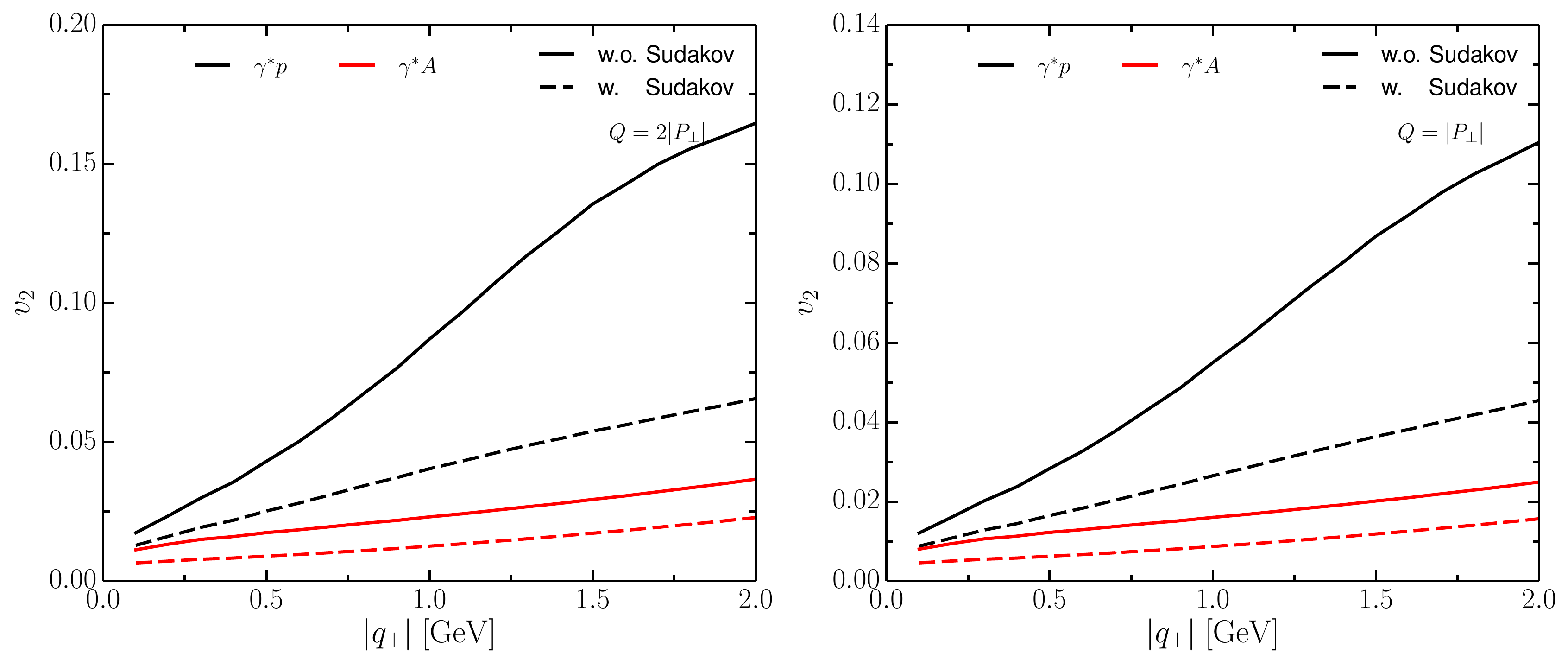}
\caption{The elliptic anisotropy $v_2$ as a function of $q_\perp$ for different scattering systems, in which $\vert P_\perp\vert=4$ GeV and $z=1/2$. The black and red curves correspond to $\gamma^\ast p$ and $\gamma^\ast A$, the solid and dashed curves denote without and with Sudakov resummations, respectively. The left panel corresponds to the case of $Q=2\vert P_\perp\vert$ and the right panel is for $Q=\vert P_\perp\vert$.}
\label{fig:fig11}
\end{figure}

In \fig{fig:fig11}, we present the results of elliptic anisotropy $v_2$ as a function of transverse momentum imbalance $q_\perp$ for different scattering systems, in which we have set $\vert P_\perp\vert=4$ GeV and $z=1/2$. The black and red curves correspond to $\gamma^\ast p$ and $\gamma^\ast A$ processes, the solid and dashed lines denote without and with Sudakov resummations, respectively. The left panel corresponds to the case of $Q=2\vert P_\perp\vert$ and the right panel is for $Q=\vert P_\perp\vert$. From this figure, for example in the left panel of \fig{fig:fig11}, one can observe that the elliptic anisotropy $v_2$ for $\gamma^\ast A$ is significantly lower than that of $\gamma^\ast p$, the main reason comes from the fact that the gluon saturation effects for heavy ion $A$ is stronger than that of proton $p$. By comparing the solid lines with dashed ones, one can see that the soft gluon resummations significantly reduces the elliptic anisotropy $v_2$ as well. Especially, in the present kinematic regime, we stress that, for $\gamma^\ast A$, the rather strong suppression of $v_2$ caused by the effects of gluon saturation and soft gluon resummations will make it very difficult to measure in experiments. On the other hand, the strong suppression for $\gamma^\ast A$ as compared to that of $\gamma^\ast p$ also as a signature of gluon saturation for heavy nucleus ion. In order to give a sizable elliptic anisotropy $v_2$ that can be measured in experiments, one possible way is to enlarge $q_\perp$ or increase $x_g$ that making $\vert q_\perp\vert/Q_{sg}$ not even small.

Although the elliptic anisotropy $v_2$ can be used to probe $xH/xG$ of the scattered target, it involves the other hard factors, such as $Q, P_\perp$, etc. Following the same manner, the polarized anisotropy can be expressed~\cite{Dumitru:2015gaa,Dumitru:2018kuw},
\begin{equation}\label{eq:v2l}
v^L_2 \equiv \frac{\int \ud \Delta\phi\cos(2\Delta\phi) \ud \sigma^L}{\int \ud \Delta\phi \ud \sigma^L}=\frac{1}{2}\frac{xH}{xG}
\end{equation}
and
\begin{equation}\label{eq:v2t}
v^T_2 \equiv \frac{\int \ud \Delta\phi\cos(2\Delta\phi) \ud \sigma^T}{\int \ud \Delta\phi \ud \sigma^T}=-\frac{1}{2}R'\frac{xH}{xG}
\end{equation}
where $R'$ is determined by kinematics in \eq{eq:17}
\begin{equation}
R'=\frac{2\left(z^2+(1-z)^2\right)P^2_\perp\epsilon^2_f-2m^2_qP^2_\perp}{\left(z^2+(1-z)^2\right)(P^4_\perp+\epsilon^4_f)+2m^2_qP^2_\perp}
\end{equation}
From \eq{eq:v2l} and \eq{eq:v2t}, one can observe that the longitudinally polarized elliptic anisotropy $v^L_2$ directly probes the ratio of $xH/xG$ and has no dependence on other additional variables, while for the transversely polarized anisotropy $v^T_2$, which involves a factor of $R'$ and has a sign change as compared to $v^L_2$. One should note that these polarized anisotropies can not be directly measured in experiments, one possible way to extract the polarized anisotropies following the relation
\begin{equation}
v_2 = \frac{R v^L_2 + v^T_2}{1+R}
\end{equation}
where $R$ is determined by comparing \eq{eq:17} and \eq{eq:18}
\begin{equation}
R = \frac{8Q^2z^3(1-z)^3 P^2_\perp}{\left(z^2+(1-z)^2\right)(P^4_\perp+\epsilon^4_f)+2m^2_qP^2_\perp}.
\end{equation}

In addition, we have also computed the longitudinally polarized anisotropy $v^L_2$ in $\gamma^\ast p$ and $\gamma^\ast A$ scattering processes, the results are very similar to that of $v_2$, therefore, we do not include it as a separated plot.

\section{Summary}\label{sec:summary}
%==============================================================================
In this paper, we have investigated the inclusive dijet cross sections and their azimuthal correlations in $ep$ and $eA$ collisions within the framework of CGC EFT. By applying the so-called ``back-to-back correlation limit'', the cross section can be expressed in terms of unpolarized and linearly polarized WW gluon distributions. Although the unpolarized and linearly polarized WW gluon distributions have little known in experiments, the novel scattering processes provide the cleanest environment to probe the two gluon TMDs, as well as allow one to quantitatively probe the gluon saturation effect inside the scattered target.

As compared to previous related studies~\cite{Dumitru:2015gaa,Dumitru:2018kuw,Marquet:2016cgx,Marquet:2017xwy}, in this work, we employed the non-linear Gaussian approximation, using the solutions of rcBK equation to give the numerical results of unpolarized and linearly polarized WW gluon distributions both in coordinate and momentum space. Results shown that the two gluon distributions have strong geometrical scaling behaviors over a broad range of transverse momentum. Especially, for the linearly polarized WW gluon distribution, it tends to be saturated and has same asymptotic behavior as the unpolarized one at large transverse momentum region, while in small transverse momentum region, in which gluon saturation effects become manifest, this distribution remains non-zero values even at small transverse momentum, and seems to be constant when transverse momentum is order of gluon saturation scale (see for example \fig{fig:fig2} and \fig{fig:fig3}). We stress that, although only the two gluon TMDs are presented in the work, the other gluon TMDs such as enter the cross sections at $pp$ and $pA$ collisions~\cite{Marquet:2016cgx,Marquet:2017xwy,Albacete:2018ruq} can also be obtained following this procedure, and hope these features depicted in the present work can be further validated by experiments in future.

Moreover, we performed the calculations of quark-antiquark pair (or dijet) cross sections and their azimuthal correlations by including the parton shower effects (or Sudakov resummations) in the scattering processes. In the calculations, we have restricted the transverse momentum $P_\perp$ and the transverse momentum imbalance $q_\perp$ of the dijet with a condition of $\vert P_\perp\vert\geq 2\vert q_\perp\vert$ to meet the ``back-to-back correlation limit''. For the cross section, although the dijet are restricted to be purely back-to-back ($\Delta\phi=\pi$), we found that the Sudakov resummations can strongly affect the transverse momentum distributions. Specifically, for $\ud\sigma/\ud\vert P_\perp\vert$, numerical results shown that the Sudakov resummations suppresses the cross sections and the suppression goes stronger with increasing $P_\perp$ (shown in \fig{fig:fig7}), while for $\ud \sigma/\ud\vert q_\perp\vert$, the Sudakov effects will significantly reduce the distributions as $q_\perp$ goes small, and seems to make the distributions to be flattened over a broad small $q_\perp$ region (\fig{fig:fig8}). As for the dijet azimuthal correlations, numerical results shown that the back-to-back correlations are significantly suppressed when the Sudakov effects are taken into account (\fig{fig:fig9}).

Furthermore, we also presented the results of dijet azimuthal correlations and elliptic anisotropy $v_2$ in the $\gamma^\ast p$ and $\gamma^\ast A$ scattering processes. By comparing the back-to-back correlations in the two scattering processes, we found that a significant suppression that caused by gluon saturation effects is visible, especially, when including the Sudakov effects, that make the back-to-back correlations to be more suppressed. We also found that similar suppression can be observed in the elliptic anisotropy $v_2$ as well. Especially, in the studied kinematic regime, we stress that, for $\gamma^\ast A$, the rather strong suppression of $v_2$ caused by the effects of gluon saturation and soft gluon resummations will make it very difficult to measure for experiments as compared to that of $\gamma^\ast p$ (see \fig{fig:fig11}). On the other hand, as compared to $\gamma^\ast p$, the rather strong suppressions of back-to-back correlations and elliptic anisotropy in $\gamma^\ast A$ is a solid signature of gluon saturation for heavy nucleus, which can be further validated or tested at a future EIC. 

We should note that, for the dijet production, although our numerical results are performed only on parton level, similar trends are expected to hold for dihadrons, this can be done by convoluting the dijet cross section with fragmentation functions. We leave it as future study.

\begin{acknowledgments}
	This work was partially supported by the Ministry of Science and Technology (MoST) under grant No. 2016YFE0104800, the Fundamental Research Funds for the Central Universities under Grant No. CCNU19ZN019 and the postdoctral science and technology project of Hubei Province under Grant No. 2018Z27. 
\end{acknowledgments}

%\bibliography{../../../refs/refs.bib}
	
\end{document}